\documentclass[onecolumn]{els-mrw}
\usepackage{amsmath,amssymb,amsfonts,amsthm,makeidx,graphicx}
\usepackage{txfonts}
\usepackage{helvet}
\usepackage[pdftex,colorlinks,breaklinks]{hyperref}
\newcommand{\re}{\mathop{\mathrm{Re}}}
\newcommand{\im}{\mathop{\mathrm{Im}}}
\newcommand{\Heff}{\mathcal{H}_{\mathrm{eff}}}
\newcommand{\aver}[1]{\left\langle{#1}\right\rangle}
\newcommand{\T}{\mathcal{T}}

\begin{document}
\chapter{Random Matrix Theory for Chaotic Wave Scattering and Transport}\label{chap1}
\author[1]{Yan V. Fyodorov}%
\author[2]{Dmitry V. Savin}%
\address[1]{\orgname{King's College London}, \orgdiv{Department of Mathematics}, \orgaddress{Strand, London, WC2R 2LS, UK}}
\address[2]{\orgname{Brunel University of London}, \orgdiv{Department of Mathematics}, \orgaddress{Uxbridge, UB8 3PH, UK}}
\articletag{Chapter Article tagline: Jun 09, 2026}

\maketitle

\begin{glossary}[Keywords]
Quantum chaos, random matrices, chaotic scattering, complex resonances, time delays, nonorthogonality
\end{glossary}

\begin{abstract}[Abstract]
We review random matrix approaches to chaotic wave scattering and transport in open systems. Starting from the effective non-Hermitian Hamiltonian formulation, we discuss the scattering matrix,  reaction matrix, time delays, and complex resonances as complementary probes of open chaotic dynamics. We emphasize universal statistics governed by symmetry, openness, and channel coupling. Topics include the maximum-entropy description of fixed-energy scattering and its applications to quantum transport, energy correlations, resonance and eigenfunction statistics, and selected wave-chaotic phenomena induced by finite absorption. The focus throughout is on non-perturbative methods and universal structures underlying   open quantum and wave chaotic systems.
\end{abstract}

\section{Introduction}\label{seec:intro}

Scattering of waves in complex systems has long been a subject of intensive study, with paradigmatic examples ranging from compound-nucleus reactions \cite{verb85,mitc10} and coherent electronic transport \cite{been97,alha00} to the propagation of electromagnetic waves in random media \cite{shi15} and experiments on microwave billiards \cite{kuhl05a}. In all these settings, the central object is the on-shell scattering matrix $S(E)$, whose matrix elements give the probability amplitudes for reflection and transmission between the channels open at energy $E$ \cite{mello_kumar2004book}. These amplitudes display a pronounced energy dependence in the vicinity of resonances, which correspond to long-lived quasibound states formed during the scattering process. Analytically, such resonances appear as complex poles $\mathcal{E}_n=E_n-\frac{i}{2}\Gamma_n$ of the scattering matrix, with $E_n$ and $\Gamma_n>0$ denoting the position and width of the $n$th resonance, respectively. In wave-chaotic systems the resonances form a dense and intrinsically fluctuating set, leading to strong variations of scattering observables as the energy or other control parameters are changed. Random matrix theory (RMT) provides a natural framework for describing the universal statistical properties of such fluctuations \cite{guhr98}.

The general RMT approach to scattering originates largely in the foundational work of Verbaarschot, Weidenm\"uller, and Zirnbauer \cite{verb85} and is based on representing the $S$-matrix in terms of an effective non-Hermitian Hamiltonian $\Heff$ of the open system. Within this formulation, the complex resonance energies $\mathcal{E}_n$ are identified with the eigenvalues of $\Heff$ \cite{soko89}. Imposing statistical assumptions on the matrix elements of the underlying Hamiltonian then yields a natural extension of the Hermitian random-matrix description of closed chaotic systems to the non-Hermitian setting appropriate for open systems \cite{soko89,fyod97}; see \cite{scho15} for a recent pedagogical review. One of the main strengths of this framework is that it treats spectral and scattering characteristics on the same footing, while remaining flexible enough to incorporate additional non-idealities such as disorder, absorption, or gain \cite{fyod05r}. Its most natural domain of applicability is the case in which the system can be clearly separated into an internal chaotic region and external scattering channels. Other classes of open disordered systems, for which such a separation is not available -- for example, resonant scattering by randomly positioned dipole scatterers embedded in a medium -- also admit random-matrix descriptions, but these require rather different formulations; see \cite{Ski11} for further discussion.

More broadly, the effective Hamiltonian approach may be viewed as a particular realisation of the general input--output theory of linear open systems, developed in system theory and engineering mathematics and going back to the seminal work of M.~Liv\v{s}ic \cite{Livsic}; for a concise account in the present context, see \cite{fyod00}. In this chapter we review selected topics in the universal statistics of resonances and scattering observables, with the main emphasis on results obtained by non-perturbative methods. The chapter may be regarded as a substantially updated and expanded version of the authors' earlier review \cite{fyod11ox}, written more than fifteen years ago. Since then, significant progress has been achieved in many of the directions covered there, and the present account is intended to reflect the most important of these developments, while inevitably being less detailed on some of the earlier results discussed in the original review. For broader treatments, including applications and experimental verifications, we refer the reader to the reviews and books cited above. We also note that many of the methods and ideas surveyed here also extend beyond the universal RMT regime: in particular, more structured random-matrix ensembles, such as banded models, can be used to incorporate effects related to Anderson localization. These developments lie beyond the scope of the present chapter; for further discussion and additional references, see an earlier review \cite{kott05} and more recent papers \cite{Gaspard2022,Fyo24,fyodmeib2025}.

\section{Scattering setup}\label{sec:setup}

Consider a wave-chaotic system coupled to the exterior through $M$ scattering channels that are open at energy $E$. The number of open channels is typically finite; for instance, in mesoscopic transport it is fixed by transverse quantization of propagating modes in the leads attached to a quantum dot. Let ${\bf a}=(a_1,\ldots,a_M)$ and ${\bf b}=(b_1,\ldots,b_M)$ denote the amplitudes of incoming and outgoing waves in these channels. By definition, the $M \times M$ scattering matrix gives the linear input--output relation for \emph{travelling-wave} amplitudes
\begin{equation}\label{S_def}
{\bf b}=S(E)\,{\bf a}\,, \qquad S^{\dagger}(E)S(E)=1\,.
\end{equation}
The second equation above expresses the unitarity condition representing the flux conservation in energy conserving systems.

In many wave settings (e.g. microwave cavities, quantum graphs),  it is also natural to work with the \emph{standing-wave} combinations ${\bf a}\pm{\bf b}$. These may be related linearly through
\begin{equation}\label{K_def}
{\bf a}-{\bf b}= iK(E)\,({\bf a}+{\bf b})\,, \qquad  K^{\dagger}(E) = K(E)\,.
\end{equation}
The matrix $K$ is known as Wigner's reaction matrix, which is a Hermitian matrix (at real $E$) for ideal systems without internal losses. Up to channel-dependent normalizations, ${\bf a}+{\bf b}$ represents the field amplitude, whereas ${\bf a}-{\bf b}$ represents the associated field current or flux. In electromagnetics these quantities may be identified with the port voltage and current, respectively, so that the matrix $iK$ relating them plays the role of an admittance (or, correspondingly, impedance $(iK)^{-1}$) matrix \cite{grad14}. Equations \eqref{S_def} and \eqref{K_def} are equivalent parameterizations of the same scattering process, expressed in travelling-wave and standing-wave variables, respectively.

\subsection{Resonances and the effective Hamiltonian}

Assuming the separation into an internal region supporting $N$ bound states and an external continuum supporting $M$ scattering channels, the coupling between the two regions is described by an $N \times M$ matrix $V$, whose columns encode the amplitudes coupling the internal states to the individual channels. These coupling amplitudes are taken to be energy-independent within the energy window of interest. This approximation is justified away from channel thresholds and when the spectrum is sufficiently dense that the couplings vary only slowly on the scale of individual resonances, as is typically the case in chaotic systems. Neglecting direct (prompt) processes, the resonant part of the $S$-matrix can then be expressed in terms of the reaction matrix as follows \cite{verb85}
\begin{equation}\label{S1}
 S(E) = \frac{1-iK(E)}{1+iK(E)} \,, \qquad K(E) = \frac{1}{2}V^{\dag}\frac{1}{E-H} V \,,
\end{equation}
where $H$ is the $N\times N$ Hermitian Hamiltonian of the corresponding closed system. For real energy $E$ and in the absence of internal losses the $K$ matrix is Hermitian, which makes the unitarity of the $S$-matrix immediate from Eq.~(\ref{S1}).

By purely algebraic transformations, the above expression can be equivalently represented in terms of the $N\times N$ effective non-Hermitian Hamiltonian $\Heff$ characterizing open counterpart of the quantum or wave chaotic system:
\begin{equation}\label{S2}
 S(E) = 1- iV^{\dag} \frac{1}{E-\Heff} V \,, \qquad \Heff = H - \textstyle\frac{i}{2}VV^{\dag}\,.
\end{equation}
The anti-Hermitian part of $\Heff$ has a factorized structure, which ensures the $S$-matrix unitarity. Physically, it accounts for decay through the open channels and transforms the real energy levels of the closed system into $N$ \emph{complex} resonances $\mathcal{E}_n=E_{n}-\frac{i}{2}\Gamma_{n}$, given by the eigenvalues of $\Heff$ and appearing as poles of $S(E)$. Since $\Heff$ is non-Hermitian, its right and left eigenvectors form a bi-orthogonal set; they define the corresponding resonance states (quasimodes) and feature in the residues of $S(E)$ at the pole positions. The pole structure of $S(E)$ implied by Eq.~\eqref{S2} provides the starting point for the statistical description of fluctuations in wave chaotic scattering.

Complementary to this stationary picture, temporal aspects of the scattering process may also be characterized in terms of the $S$-matrix. In particular, the widths $\Gamma_n$ determine the finite lifetimes ($\sim\hbar/\Gamma_n$) of quasibound states formed at the intermediate stage of the scattering event. Experimentally, they can be accessed through spectroscopic analysis of relevant decay spectra \cite{kuhl08,difa12}, see also Chapter X in this volume \cite{Kolovsky_chapter}. More generally, however, the temporal response is encoded in the variation of $S$ with energy, leading to the notion of delay times and related observables. The central role here is played by the $M\times M$ Hermitian matrix $Q(E)=-i\hbar\,S^\dagger dS/dE$, known as the Wigner-Smith time-delay matrix, see \cite{fyod97,deCar02} for the introduction and historical background and \cite{kolo13,texier2016wigner} for practical applications.

Within the resonance approximation, the energy dependence enters only through the Green function $(E-\Heff)^{-1}$ describing propagation in the open system governed by the effective Hamiltonian $\Heff$. The energy derivative of the $S$-matrix is then straightforward to evaluate, yielding the following convenient representation for the Wigner–Smith matrix \cite{soko97} (we set $\hbar=1$ henceforth):
\begin{equation}\label{Q}
 Q(E) = \Psi^\dag(E)\Psi(E)\,, \qquad \Psi(E) = \frac{1}{E-\Heff} V\,.
\end{equation}
The $a$th column $\Psi_a$ of the $N\times M$ matrix may be interpreted \cite{soko97} as the internal part of the scattering wave function generated by incidence in channel $a$ at energy $E$. Accordingly, the matrix element $Q_{ab}(E)=\Psi_a^\dagger(E)\Psi_b(E)$ measures the overlap of the internal parts corresponding to incidence in channels $a$ and $b$. In particular, the norm of $\Psi_a$ gives the diagonal element $Q_{aa}$, which therefore has the meaning of a channel-resolved mean delay time. Averaging over the channels yields the so-called Wigner time delay $\tau_W=M^{-1}\mathrm{Tr}\,Q$. It can be equivalently expressed through the energy derivative of the total scattering phase $\Phi(E)=-i\ln\det S(E)$ as $\tau_W=M^{-1}d\Phi(E)/dE$.

Within the same approximation, the only singularities of the scattering matrix are its poles $\mathcal{E}_n=E_n-\frac{i}{2}\Gamma_n$. For a unitary scattering matrix, the corresponding zeros are located at the complex-conjugate points $\mathcal{E}_n^*$. These two properties imply the general representation
\begin{equation}\label{dets1}
 \det S(E)=\prod_{k=1}^N \frac{E-\mathcal{E}_k^*}{E-\mathcal{E}_k}\,,
\end{equation}
which also follows directly from Eq.~\eqref{S2}, since $\det S(E)=\frac{\det(E-\Heff^\dagger)}{\det(E-\Heff)}$. As a consequence, both the total scattering phase and time-delay characteristics can be expressed in terms of the resonance poles alone. In particular, Eq.~\eqref{dets1} further implies that the Wigner time delay is entirely determined by the resonance poles \cite{lehm95b}:
\begin{equation}\label{tau_W}
 \tau_W(E) = \frac{1}{M}\mathrm{Tr}\,Q
 = \frac{1}{M}\sum_{n=1}^N\frac{\Gamma_n}{(E-E_n)^2+\Gamma_n^2/4}\,.
\end{equation}
This expression is valid at arbitrary degree of the resonance overlap. It also shows that the Wigner time delay (\ref{tau_W}) can be treated as the density of states of the open system. In particular, averaging over a narrow energy window gives $\overline{\tau}_W = 2\pi/\Delta M$, where $\Delta$ is the mean level spacing around $E$. The associated fundamental timescale is the Heisenberg time $t_H = 2\pi/\Delta$, so that $\overline{\tau}_W=t_H/M$.

\subsection{Statistical assumptions}

Universal statistical properties of resonances and of fluctuations in chaotic scattering are obtained by replacing the Hamiltonian of the closed counterpart of the system by a random-matrix model of the appropriate symmetry class. Conventionally, the Hermitian part $H$ is taken from the \emph{Gaussian Orthogonal Ensemble} (GOE, Dyson index $\beta=1$) or the \emph{Gaussian Unitary Ensemble} (GUE, $\beta=2$), depending on whether time-reversal invariance (TRI) is preserved or broken, respectively. The \emph{Gaussian Symplectic Ensemble} (GSE, $\beta=4$) becomes relevant for TRI systems in which additional symmetries associated with spin degrees of freedom must be retained. In the limit $N\to\infty$, local spectral correlations on the scale of the mean level spacing $\Delta\sim\frac{1}{N}$ become universal, i.e. independent of microscopic details such as the precise probability distribution of $H$ or the slow energy dependence of the mean density of states. Without loss of generality, one may therefore restrict attention to the spectral centre. In the conventional normalization of the Gaussian ensembles, the large-$N$ mean eigenvalue density then follows the Wigner semicircle law on $[-2,2]$, and the corresponding mean level spacing at $E=0$ is $\Delta=\pi/N$. Results away from the centre are recovered by appropriate rescaling with the local value of $\Delta$. Further details on RMT techniques are given in Chapter X~\cite{Kieburg_chapter}.

A similar universality holds for the coupling amplitudes $V_n^c$, provided the number of open channels remains finite as $N\to\infty$, or more generally satisfies $M\ll N$~\cite{lehm95a}. In this regime, the detailed statistical assumptions about the couplings are immaterial: the amplitudes may be taken either as fixed \cite{verb85} or as random \cite{soko89} variables, and enter final expressions only through the transmission coefficients
\begin{equation}\label{T}
 T_c\equiv1-|\overline{S}_{cc}|^2=\frac{4\kappa_c}{(1+\kappa_c)^2}\,, \qquad
 \kappa_c=\frac{\pi\|V^c\|^2}{2N\Delta}\,.
\end{equation}
Here $\overline{S}$ denotes the average (optical) scattering matrix. In the absence of direct processes, one may choose the channel basis so that $\overline{S}$ is diagonal: $\overline{S}_{ab}=0$ ($a\neq b$) excludes prompt transitions between different channels, while the diagonal elements $\overline{S}_{cc}=\frac{1-\kappa_c}{1+\kappa_c}$ describe direct reflection due to imperfect coupling. The set $\{T_c\}$ thus provides the natural input parameters characterizing the strength of coupling to the continuum: $T_c=1$ corresponds to perfect coupling in channel $c$, whereas $T_c\ll1$ describes weak coupling and an almost closed channel.

\section{Fixed-energy scattering statistics}\label{distr}

At fixed real scattering energy $E$ and in the flux-conserving limit, the statistical properties of transport observables are fully encoded in the distribution of the unitary matrix $S=S(E)$. A conceptually simple alternative to the Hamiltonian approach is to characterize this distribution directly in channel space, without explicit reference to the underlying closed-system Hamiltonian. In this formulation, all system-specific information is contained in the average (optical) scattering matrix $\overline{S}$, which accounts for prompt processes and non-ideal coupling. Under the maximum-entropy assumption, the probability density of $S$ is then uniquely determined by $\overline{S}$ and is given by the Poisson kernel \cite{mello_kumar2004book}:
\begin{equation}\label{poisson}
 P_{\overline{S}}(S)  =\frac{1}{V_\beta}\left| \frac{\det[1- \overline{S}^{\dag}\overline{S}]}{
\det[1- \overline{S}^{\dag}S]^2}\right|^{(\beta M +2-\beta)/2},
\end{equation}
where $V_\beta$ is a normalization constant. Although this expression also follows from the Hamiltonian formulation, Eq.~\eqref{S2}, in the limit $N\to\infty$ \cite{brou95}, it can be derived directly from statistical assumptions imposed on $S$ itself, as was first done by Mello and co-workers in the mid-1980's. Below we briefly outline this maximum entropy approach; for historical remarks and a more detailed exposition, see~\cite{mello_kumar2004book}.

\subsection{Maximum entropy approach}\label{sec:max_entr}

The maximum entropy approach treats $S$ as the fundamental random object and seeks the least biased probability distribution $P_{\overline{S}}(S)$ compatible with the known information, once the symmetry class and the optical scattering matrix $\overline{S}$ have been specified. To this end, one introduces the information entropy $\mathcal{S}=-\int d\mu(S)\,P_{\overline{S}}(S)\ln P_{\overline{S}}(S)$, where $d\mu(S)$ is the invariant (Haar) measure on the corresponding unitary matrix space, supplemented by the appropriate symmetry constraints: unitary $S$ is also symmetric for $\beta=1$ or self-dual for $\beta=4$.  The essential physical input comes from analyticity and ergodicity. Causality implies that $S(E)$ is analytic in the upper half of the complex energy plane, while the assumed stationarity and ergodicity identify ensemble averages with spectral averages. Together these properties yield the \emph{analyticity-ergodicity} requirement $\aver{S^k}\equiv\int d\mu(S)\,S^k P_{\overline{S}}(S)=\overline{S}^{\,k}$ for all $k\geq1$. Maximizing $\mathcal{S}$ subject to these constraints gives the Poisson kernel~\eqref{poisson}, with $V_\beta=\int d\mu(S)$ being the volume of the matrix space. As a result, this distribution is characterized by the reproducing property $\int d\mu(S)\,f(S)\,P_{\overline{S}}(S)=f(\overline{S})$ for analytic $f$, showing that all non-universal information is encoded in $\overline{S}$ alone.

The case $\overline{S}=0$, corresponding to ideal (perfect) coupling, is of particular importance. In physical terms, it describes the absence of prompt or direct processes, so that scattering is entirely due to the equilibrated chaotic dynamics. The $S$-matrix is then distributed uniformly over the appropriate Dyson circular ensemble, i.e. $P_0(S)=\mathrm{const}$ with respect to the invariant Haar measure. Statistical averaging therefore reduces to integration over the corresponding unitary group \cite{mello_kumar2004book,brou96}. For $\overline{S}\neq0$, the presence of non-ideal coupling and/or direct processes makes the problem analytically more difficult. It may nevertheless be reduced to the ideal-coupling case through
\begin{equation}\label{pois-map}
 S_0 = (t'_1)^{-1} (S-\overline{S})(1-\overline{S}^{\dag}S)^{-1} t_1^{\dag}\,,
\end{equation}
where $S_0$ belongs to the appropriate circular ensemble and $t_1,t_1'$ are sub-blocks of a $2M\times2M$ unitary matrix $S_1={r_1\  t_1' \choose t_1 \ r_1'}$ of the same symmetry class. The Jacobian of this transformation yields precisely the Poisson kernel \eqref{poisson}, upon identifying $r_1=\overline{S}$. Physically, this corresponds to decomposing scattering into a prompt stage $S_1$ and an equilibrated stage $S_0$. Equation~\eqref{pois-map} then expresses the composition of these two stages and provides a convenient mapping from the problem with direct processes to the corresponding one without them. A number of useful applications of this construction were discussed in \cite{gopa98,savi01,fyod05r}.

In quantum transport applications, it is natural to partition the scattering channels according to the propagating modes supported by the attached leads. In the standard two-terminal geometry, one distinguishes $N_1$ channels in the `left' lead and $N_2$ channels in the `right' lead, so that the total number of channels is $M=N_1+N_2$. The scattering matrix then takes the block form
\begin{equation}\label{S}
  S = \left( \begin{array}{cc} r & t' \\ t & r' \end{array}\right)
    = \left( \begin{array}{cc} u & 0 \\ 0 & v \end{array}\right)
    \left( \begin{array}{cc} -\sqrt{1{-}\hat\tau^T\hat\tau} & \hat\tau^T \\
    \hat\tau & \sqrt{1{-}\hat\tau\hat\tau^T} \end{array}\right)
    \left( \begin{array}{cc} u' & 0 \\ 0 & v' \end{array}\right),
\end{equation}
where $r$ ($r'$) is a $N_1\times N_1$ ($N_2\times N_2$) reflection matrix in the left (right) lead and $t$ ($t'$) is a rectangular matrix of transmission from left to right (right to left). Unitarity implies that the four Hermitian matrices $tt^{\dag}$, $t't'^{\dag}$, $1-rr^{\dag}$ and $1-r'r'^{\dag}$ share the same set of $n\equiv\mathrm{min}(N_1,N_2)$ eigenvalues \{$\T_k\in[0,1]$\}, known as the \emph{transmission eigenvalues}. The second equality in~\eqref{S} is the polar decomposition of $S$, obtained from the singular-value decompositions $t=v\hat\tau u'$ and $t'=u\hat\tau^{T}v'$. Here $\hat\tau$ is an $N_1\times N_2$ rectangular diagonal matrix with nonzero entries $\hat\tau_{kk}=\sqrt{\T_k}$, $k=1,\ldots,n$. The matrices $u$ and $u'$ ($v$ and $v'$) are $N_1\times N_1$ ($N_2\times N_2$) unitary matrices (note that $u^*u'=v^*v'=1$ for $\beta=1,4$).

When $S$ is drawn from one of Dyson's circular ensembles (the case of ideal coupling), the angular matrices are Haar distributed, whereas the nontrivial statistics resides entirely in the transmission eigenvalues. Their joint probability density function (JPDF) is \cite{been97} 
\begin{equation}\label{jpdf}
  \mathcal{P}_\beta(\{\T_i\}) = \frac{1}{\mathcal{N}_\beta} |\Delta(\mathcal{T})|^{\beta}\prod_{i=1}^n
    \T_i^{\alpha -1}\,, \qquad \alpha\equiv\frac{\beta}{2}(|N_1-N_2|+1)\,.
\end{equation}
Here $\mathcal{N}_\beta$ is a normalization constant, and throughout we use the standard notation $\Delta(x)=\prod_{i<j}(x_i-x_j)$ for the Vandermonde determinant in the variables $\{x_i\}$. The density~\eqref{jpdf} specifies the so-called \emph{Jacobi Ensemble} in RMT \cite{Forrester}, which therefore plays a central role in this context. It thus provides the natural starting point for the statistical analysis of transport through wave chaotic systems; see \cite{been97} for a review.

\subsection{Transport moments and the Selberg integral}

Within the Landauer--B\"uttiker formalism many transport observables can be expressed in terms of the transmission eigenvalues $\{\mathcal{T}_k\}$ \cite{blan00r}. Important examples are the conductance $g=\mathrm{tr}(tt^{\dag})=\sum_{i=1}^{n}\T_i$ and shot-noise power $p=\sum_{i=1}^{n}\T_i(1-\T_i)$ (at zero temperature), both measured in their natural units. Since these quantities are linear statistics of the transmission eigenvalues, their mean values may be obtained by averaging over the transmission block $t$ or, equivalently, from the mean eigenvalue density~\cite{arau98}. The latter can be derived by exploiting the connection with the Jacobi ensemble \cite{vivo08a}. By contrast, higher moments involve products of $\T_j$'s and therefore require integration over the full JPDF. This becomes possible through the profound connection between Eq.~\eqref{jpdf} and the Selberg integral, first pointed out in \cite{savi06}.

The Selberg integral is defined by the following expression (yielding also the normalization in (\ref{jpdf}) when $a=\alpha$, $b=1$ and $c=\frac{\beta}{2}$):
\begin{equation}\label{Selberg}
 \int_{0}^{1}\!\mathrm{d}x_1\!\cdots\!\int_{0}^{1}\!\mathrm{d}x_n\,
 |\Delta(x)|^{2c} \prod_{i=1}^{n} x_i^{a-1}(1-x_i)^{b-1}=
 \prod_{j=0}^{n-1}\frac{\Gamma(1+c+jc)\Gamma(a+jc)\Gamma(b+jc)}{
 \Gamma(1+c)\Gamma(a+b+(n+j-1)c)}\,,
\end{equation}
where $\Gamma(x)$ is the gamma function. This expression is valid for integer $n\ge1$,  complex $a$ and $b$ with positive real parts, and complex $c$ with $\mathrm{Re\,}c>-\mathrm{min}[\frac{1}{n},\mathrm{Re\,}\frac{a}{n-1},\mathrm{Re\,}\frac{b}{n-1}]$ and can be considered as a multidimensional generalization of Euler's beta-function~\cite[Chapter 4]{Forrester}. The special structure of the integrand makes it possible to derive a closed set of algebraic relations for the moments $\langle\mathcal{T}_1^{\lambda_1}\cdots\mathcal{T}_m^{\lambda_m}\rangle$ of total order $r=\sum_i \lambda_i$, with integer parts $\lambda_i\geq0$. The computation of moments is thereby reduced to  solving the corresponding algebraic system~\cite{savi08}.

As a simple illustration, we consider the second-order moments that are defined by the recursion relations
$ [\alpha+2+\beta(n-1)] \langle{\T_1^2}\rangle = [\alpha+1+\beta(n-1)]\langle{\T_1}\rangle - \frac{\beta}{2}(n-1)\langle{\T_1\T_2}\rangle
$
and
$\langle{\T_1\cdots\T_m}\rangle = \prod_{j=1}^m \frac{\alpha+\beta(n-j)/2}{\alpha+1+\beta(2n-j-1)/2}
$
at $m=1,2$. Together, these expressions lead directly to the exact results for the mean and variance of the conductance, well known from \cite{been97}, as well as for the mean shot-noise power~\cite{savi06}:
\begin{equation}
 \langle{g}\rangle = \frac{N_1N_2}{M+1-\frac{2}{\beta}}, \qquad
 \mathrm{var}(g) = \langle{g}\rangle
  \frac{2(N_1-1+\frac{2}{\beta})(N_2-1+\frac{2}{\beta}) }{     \beta(M-2+\frac{2}{\beta})(M-1+\frac{2}{\beta})(M-1+\frac{4}{\beta})}\,, \qquad
 \langle{p}\rangle = \langle{g}\rangle \frac{(N_1-1+\frac{2}{\beta})(N_2-1+\frac{2}{\beta}) }{
    (M-2+\frac{2}{\beta})(M-1+\frac{4}{\beta})}\,.
\end{equation}
We note an exact relation $\mathrm{var}(g)=2\langle g\rangle\langle p\rangle/\beta N_1N_2$. In particular, it relates the Fano factor $\frac{\langle p\rangle}{\langle g\rangle}$, which quantifies the suppression of shot noise relative to Poissonian noise, to $\mathrm{var}(g)$ and hence to universal conductance fluctuations. The same method also yields explicit expressions for the conductance skewness and kurtosis, as well as for the shot-noise variance \cite{savi08}. It can be extended to the full counting statistics of charge transfer (related to moments of the type $\langle{T_1^k}\rangle$) \cite{nova07}. It is worth emphasizing that the symmetry index $\beta$ enters the Selberg integral as a continuous parameter. As a result, the method treats all three symmetry classes on the same footing and provides a powerful non-perturbative alternative to approaches based on orthogonal polynomials \cite{arau98,vivo08a}, group integration \cite{brou96,bulg06c}, and related developments~\cite{mezz11,mezz12}.

It is further possible to combine the Selberg-integral approach with the theory of symmetric functions \cite{nova08}, thereby developing a systematic framework for computing moments of the conductance and shot noise (including their mixed moments) of arbitrary order and for any number of open channels \cite{khor09}. The key idea is to expand powers of $g$ and $p$ (or more general linear statistics of the transmission eigenvalues) in the basis of Schur functions $s_\lambda$. These are symmetric polynomials in the variables $\mathcal{T}_1,\ldots,\mathcal{T}_n$, labelled by partitions $\lambda$,\footnote{A partition is a finite sequence $\lambda=(\lambda_1,\lambda_2,\ldots,\lambda_m)$ of non-negative integers, called parts, arranged in non-increasing order $\lambda_1\geq\lambda_2\geq\cdots\geq\lambda_m\geq0$. Its weight is $|\lambda|=\sum_j\lambda_j$, and its length $l(\lambda)$ is the number of non-zero parts.
}
and defined by $s_{\lambda}(\mathcal{T})
=\det\{\mathcal{T}_i^{\lambda_j+n-j}\}_{i,j=1}^n/\det\{\mathcal{T}_i^{\,n-j}\}_{i,j=1}^n$.
From the viewpoint of representation theory, Schur functions are the irreducible characters of the unitary group and therefore form an orthogonal family. This orthogonality can be exploited to determine the coefficients in the Schur function expansion by integration over the unitary group. The general form of such an expansion reads:
\begin{equation}\label{exp}
  \prod_{i=1}^n \biggl(\sum_{j=-\infty}^{+\infty} a_j{\T_i}^j \biggr)
  = \sum_{\lambda}c_{\lambda}(a) s_{\lambda}(\T),
  \qquad
  c_{\lambda}(a) \equiv \det\bigl\{a_{\lambda_k-k+l}\bigr\}_{k,l=1}^n.\quad
\end{equation}
The summation extends over all partitions $\lambda$ of length at most $n$, including the empty partition for which $s_{\lambda}=1$. The averages of Schur functions with respect to the JPDF \eqref{jpdf} can then be computed using integration formulas due to Hua \cite{Hua}. In this way one finds
\begin{equation}\label{mg1aa}
 \langle s_{\lambda} \rangle_{\beta=1} = c_{\lambda}\ \prod_{j=1}^{l(\lambda)} \frac{(\lambda_j+N_1-j)!}{(N_1-j)!}\ \frac{(\lambda_j+N_2-j)!}{(N_2-j)!}
 \frac{(M-l(\lambda)-j)!}{(\lambda_j+M-l(\lambda)-j)!}
 \prod_{1\le i\le j\le l(\lambda)} \frac{M+1-i-j}{M+1+\lambda_i+\lambda_j-i-j} \,,
\end{equation}
where we have introduced the coefficient
$
c_{\lambda} = \prod_{1\le i<j\le l(\lambda)}(\lambda_i-i-\lambda_j+j)/\prod_{j=1}^{l(\lambda)}\, (l(\lambda)+\lambda_j-j)!
$,
and
\begin{equation}\label{mg2a}
 \langle s_{\lambda} \rangle_{\beta=2} = c_{\lambda} \prod_{j=1}^{l(\lambda)} \frac{(\lambda_j+N_1-j)!}{(N_1-j)!} \frac{(\lambda_j+N_2-j)!}{(N_2-j)!} \frac{(M-j)!}{(\lambda_j+M-j)!}\,,\quad
\end{equation}
for the orthogonal ($\beta=1$) and unitary ($\beta=2$) symmetry classes, respectively.

As an application of the method, we consider the moments of $\epsilon g+p$. This quantity has a physical meaning of the total noise including both thermal and shot-noise contributions \cite{blan00r}. Using expansion (\ref{exp}), one finds~\cite{khor09} that the $r$-th moment of the total noise is given by
\begin{equation}\label{gpexp}
 \langle(\epsilon g + p)^r \rangle = r! \sum_{m=0}^{r} (-1)^m (1+\epsilon)^{r-m}
  \!\!\!\!\sum_{|\lambda|=r+m} f_{\lambda,m} \langle s_{\lambda} \rangle,\ \
\end{equation}
where the inner sum runs over all partitions of weight $r+m$. The coefficients $f_{\lambda,m}$ are given by $f_{\lambda,m}=\sum \det\{[k_i!(\lambda_i-i+j-2k_i)!]^{-1}\}$, where the integers $k_i$ range from $0$ to $m$ subject to the constraint $k_1+\cdots+k_{l(\lambda)}=m$. In the limit $\epsilon\to\infty$, Eq.~\eqref{gpexp} reduces to the moments of the conductance,
$\langle g^r\rangle=r!\sum_{|\lambda|=r} c_\lambda \langle s_\lambda\rangle$,
since $f_{\lambda,0}=c_\lambda$; for $\beta=2$ these were first obtained in \cite{nova08} (see also \cite{osip08} for an alternative derivation). In the opposite limit $\epsilon=0$, one recovers all moments of the shot noise and hence, by standard recursion relations, all of its cumulants.

Beyond yielding explicit low-order results, the method is also well suited to the symbolic computation of higher-order cumulants with the aid of computer algebra. This makes it possible to analyse their asymptotic behaviour in the regime $N_{1,2}\gg1$ and to extract the leading terms in the large-$n$ expansion. In the important case of symmetric leads, $N_1=N_2=n$, the leading behaviour of the conductance and shot-noise cumulants was conjectured in \cite{khor09} on the basis of exact finite-$n$ calculations and takes the form ($r\geq3$)
\begin{equation}
 \langle\langle g^r \rangle\rangle \simeq \frac{(r-1)!}{4(2\beta n)^r} \times \left\{
 \begin{array}{ll}
  1, & \mbox{$\beta=1$: odd $r$} \\[1ex]
  \displaystyle \frac{-2n(r{-}3)!!}{r!!}, & \mbox{$\beta=1$: even $r$} \\[1ex]
  1, & \mbox{$\beta=2$: even $r$ only}
 \end{array}\right.,
 \qquad
 \langle\langle p^r \rangle\rangle \simeq \frac{(r-1)!}{8(4\beta n)^r} \times \left\{
 \begin{array}{ll}
  1, & \mbox{$\beta=1$: odd $r$} \\[1ex]
  \displaystyle \frac{-4n(r{-}3)!!}{r!!}, & \mbox{$\beta=1$: even $r$} \\[1ex]
  1, & \mbox{$\beta=2$: all $r$}
 \end{array}\right. .
\end{equation}
For $\beta=2$, all odd conductance cumulants vanish identically in the case of symmetric leads. This follows from the symmetry of the JPDF~\eqref{jpdf} at $\alpha=1$, which implies that the conductance distribution is symmetric about $n/2$ \cite{savi08}. More generally, these observations point to a deeper integrable structure underlying transport statistics. This structure was first identified for the unitary symmetry class in \cite{osip08,osip09}, where the relevant generating functions were embedded into the framework of certain matrix integrals and the associated integrable hierarchies, providing an alternative non-perturbative derivation of the $\beta=2$ cumulants and their large-$n$ asymptotics. The integrable-theory approach was subsequently extended to the orthogonal ($\beta=1$) and symplectic ($\beta=4$) symmetry classes in \cite{mezz13}. Within this framework, the cumulant generating functions for conductance and shot noise satisfy nonlinear differential equations, while the corresponding cumulants obey non-perturbative recurrence relations that enable efficient computation at arbitrary order. In particular, this extension confirms the conjectured large-$n$ behaviour in the remaining symmetry classes and provides a controlled scheme for deriving higher-order, or $1/n$, corrections.

\subsection{Conductance and shot-noise distributions}

Complementary to the study of moments and cumulants, one may consider the full probability density of transport observables. By definition, such distributions are obtained by projecting the joint density of transmission eigenvalues onto the relevant linear statistics. In the case of the conductance, one has
$P(g)=\left\langle \delta\!\left(g-\sum_{i=1}^{n}\mathcal{T}_i\right)\right\rangle$,
where the average is taken over the JPDF~\eqref{jpdf}. For small numbers of channels this integration can be carried out explicitly. In particular, exact expressions are known for $n=1,2$ and arbitrary $N_2\equiv L\geq N_1\equiv n$~\cite{khor09}:
\begin{equation}\label{prob_g}
 P(g) = \left\{
\begin{array}{ll}
  (\beta L/2)g^{\beta L/2-1} \quad (0<g<1), & n=1 \\[1ex]
  Lg^{\beta L-1} [ X_1 - (-1)^{(\beta L-1)/2} X_2\, \Theta(g-1) \sum_{j=0}^{\beta} {\beta\choose j} B_{1-g} (j+\frac{\beta}{2}(L-1), 1-\beta L)] \quad (0<g<2), &  n=2
 \end{array}\right. .
\end{equation}
Here the constants $X_1 = \frac{\Gamma(\beta(L+1)/2+1) \Gamma(\beta L/2) }{ \Gamma(\beta/2)\Gamma(\beta L)}$ and $X_2 = \frac{\Gamma(\beta(L+1)/2+1) }{ \Gamma(\beta)\Gamma(\beta(L-1)/2)}$, and $B_{z}(a,b)$ stands for the incomplete beta-function. These elementary cases already illustrate two generic features of the distribution: its compact support, $0\leq g\leq n$, and its sensitivity to both the symmetry index $\beta$ and the channel asymmetry $|N_1-N_2|$. Similarly, the short-noise distribution is nonzero only for $0\leq p\leq n/4$.

For larger channel numbers, direct integration over the simplex constrained by $\sum_i\mathcal{T}_i=g$ rapidly becomes cumbersome. A more efficient route is to consider integral transforms of the distribution, such as its Fourier or Laplace transform. These transforms can be expressed in compact algebraic forms: as determinants in the unitary case ($\beta=2$), and as Pfaffians in the orthogonal and symplectic cases ($\beta=1,4$). Such representations provide a practical starting point for deriving conductance or short-noise distributions at low or moderate values of $N_{1,2}$. Detailed formulae and examples were obtained using a Fourier representation in \cite{khor09} and, in a related Laplace-transform formulation covering all three invariant symmetry classes as well as crossover ensembles, in \cite{Kumar2010}.

As the number of channels increases, the distribution of the conductance, or of any linear statistics in general, approaches a Gaussian form in the bulk, reflecting the suppression of higher cumulants in the large-channel limit \cite{been97}. The exact finite-$N_{1,2}$ cumulants discussed above allow one to go beyond this leading approximation. In particular, systematic corrections to the Gaussian law can be constructed using the Edgeworth expansion in terms of the skewness, kurtosis, and higher cumulants. Such approximations based of the first four cumulants were shown to reproduce the bulk of the exact distribution with good accuracy even for relatively small numbers of channels \cite{khor09}.

The far tails and edge behaviour require a different asymptotic description. In the limit $n\to\infty$, with the channel ratio kept fixed, the distributions of linear statistics can be analysed by Coulomb-gas methods \cite{vivo08}. The transmission eigenvalues are then viewed as charges confined to $[0,1]$, while fixing the conductance or shot-noise power imposes a constraint on their equilibrium density. The resulting large-deviation analysis yields a central Gaussian regime matched to non-Gaussian tails. Near the edges of the finite support, the distributions exhibit power-law behaviour \cite{khor09}, and the crossover between bulk and tail regimes is associated with a third-order phase transition in the constrained Coulomb gas \cite{vivo08,Cunden2015}. This provides a complementary large-$n$ perspective on the exact finite-channel results and clarifies the physical origin of non-Gaussian features of linear statistics in the universal large-$n$ limit.

Finally, we mention an alternative exact approach to compute $P(g)$, which is based on recursion schemes for matrix integrals associated with the \emph{Wishart--Laguerre} ensemble. This ensemble also plays a central role in the RMT description of time delays discussed below. By relating the conductance to the largest Wishart--Laguerre eigenvalue, this method yields recurrence relations and differential equations for $P(g)$ for arbitrary channel numbers and symmetry index $\beta$~\cite{Forrester2019}, providing an efficient complement to the approaches discussed above.

\subsection{Nonideal coupling}

We have so far focused on ideal coupling, for which the $S$ matrix is uniformly distributed over the appropriate circular ensemble and the transmission eigenvalues form the Jacobi ensemble. For non-ideal coupling, the starting point is instead the Poisson kernel with nonzero optical matrix $\overline{S}$. Although the two cases are related through \eqref{pois-map}, the presence of direct processes substantially complicates the analysis: the induced distribution of transmission eigenvalues is no longer given by the simple Jacobi weight, and the determinant/Pfaffian and Selberg-integral methods described above do not apply in their original form. Nevertheless, important progress has been made in special cases. For unitary symmetry ($\beta=2$), the case of one ideal and one non-ideal lead can be reformulated in terms of reflection eigenvalues, whose JPDF was derived for chaotic cavities coupled through both ballistic and tunnel point contacts \cite{Vidal2012}. This result was used to compute densities and correlation functions of reflection eigenvalues and to analyse conductance fluctuations in asymmetric cavities.

Further developments extended the treatment of non-ideal leads to all three Dyson symmetry classes, expressing the reflection-eigenvalue JPDF in terms of hypergeometric functions of matrix arguments and revealing determinantal or Pfaffian structures in special cases \cite{Jarosz2015}. Close to the ideal-coupling limit, perturbative approaches are also available: for $\beta=2$ with one weakly non-ideal lead, symmetric-function and generalized Selberg-integral methods yield systematic expansions in the barrier transparencies for the conductance mean and variance and for the average shot noise \cite{Rodriguez2013}. Thus, although non-ideal coupling destroys the simple Jacobi-ensemble structure, a combination of Poisson-kernel methods and perturbative expansions still provides explicit transport statistics in several important regimes.

\subsection{Distribution of the time-delay matrix}

As is clear from its definition, the Wigner--Smith matrix $Q(E)$ at a given energy is intrinsically a two-point object: it involves the energy derivative and therefore probes correlations of $S(E)$ at nearby energies. Its distribution cannot be inferred from the fixed-energy distribution of $S$ alone, as in the preceding discussion, but requires information about the energy-dependent scattering process $S(E)$. The RMT solution of this problem was first obtained for ideal coupling, $\overline{S}=0$, by Brouwer, Frahm, and Beenakker~\cite{brou97a,brou99}, who determined the joint distribution of $S$ and $Q$ in the circular-ensemble setting. The generalisation to arbitrary coupling was developed more recently in \cite{Grabsch2018}.

Before discussing these results, it is useful to distinguish several related time-delay characteristics. Since $Q$ is Hermitian, it has real eigenvalues $\tau_1,\ldots,\tau_M$, the so-called \emph{proper} delay times. The diagonal elements $Q_{cc}$ are also real and represent the mean delay associated with incidence in a given scattering channel. A third set of quantities is provided by the \emph{partial} delay times $\tilde{\tau}_c=d\phi_c/dE$, defined as the energy derivatives of the scattering eigenphases $\phi_c$, $c=1,\ldots,M$. Their exact density is known for arbitrary $M$ and coupling \cite{fyod97,fyod97a}. The key difference between the proper and partial delay times lies in the order in which the diagonalization and energy differentiation of $S(E)$ are performed. This leads to distinct statistical properties \cite{brou99}.

The relation between these quantities can be clarified by using the spectral decomposition for $S=U_S e^{i\hat{\phi}} U_S^\dagger$. One finds \cite{savi01}
\begin{equation}
  Q = U_S \frac{d\hat{\phi}}{dE} U_S^\dagger
  + iS^\dagger\left[U_S\frac{dU_S^\dagger}{dE},S\right],
\end{equation}
where $\hat{\phi}=\mathrm{diag}(\phi_1,\ldots,\phi_M)$ and $[\cdot,\cdot]$ denotes a commutator. The first term involves the partial delay times. Clearly, they are simply given by the diagonal elements of $Q$ in the eigenbasis of $S$. The second term accounts for the energy dependence of the scattering eigenvectors (the columns of $U_S$) and mixes different eigenchannels. It is therefore this term that distinguishes the proper delay times, i.e. the eigenvalues of the full matrix $Q$, from the partial delay times. Despite these differences, all three time-delay sets satisfy the same sum rule:
\begin{equation}\label{td_sumrule}
  \frac{1}{M}\sum_{c=1}^M \tau_c  = \frac{1}{M}\sum_{c=1}^M \tilde{\tau}_c
  = \frac{1}{M}\sum_{c=1}^M Q_{cc} = \frac{1}{M}\mathrm{Tr}\,Q  = \tau_W .
\end{equation}
For statistically equivalent channels, this identity implies the equality of the average delay times:
$\aver{\tau_c} = \aver{\tilde{\tau}_c} = \aver{Q_{cc}} = \aver{\tau_W} = t_H/M$.

\subsubsection{Perfect coupling}

For $\overline{S}=0$, the scattering matrix at fixed energy is uniformly distributed in the appropriate circular ensemble. The key step in \cite{brou97a} was to extend this invariance to the distribution functional of $S(E)$. Namely, the whole energy-dependent $S$-matrix ensemble of a chaotic system is invariant under a transformation $S(E)\to U S(E) U'$, with the unitary energy-independent matrices $U$ and $U'$ constrained by the relevant symmetry. A proof of this invariance property, given in \cite{brou99}, uses the Hamiltonian formulation of chaotic scattering and the fact that, in the limit $N\to\infty$, the statistical properties of $S(E)$ become independent of the detailed random-matrix model chosen for the closed-system Hamiltonian $H$. As a consequence, the proper delay times turn out to be statistically independent of $S$ and are governed by a Laguerre ensemble after inversion. More precisely, introducing the scaled rates (inverse proper delay times) $\gamma_a=t_H/\tau_a$, their JPDF is
\begin{equation}\label{jpdf_td}
 {\cal P}(\gamma_1,\ldots,\gamma_M)\propto \prod_{a<b}|\gamma_a-\gamma_b|^{\beta}
 \prod_{a=1}^M \gamma_a^{\beta M/2} e^{-\beta\gamma_a/2}.
\end{equation}
The density~\eqref{jpdf_td} specifies the so-called \emph{generalized Laguerre (or Wishart–Laguerre) ensemble} \cite{Forrester}. This result provides one of the clearest links between chaotic scattering and the classical random-matrix ensembles beyond Dyson's circular ensembles.

A useful note is that one should distinguish between the conventional Wigner--Smith matrix, $Q=-iS^\dagger dS/dE$ and its symmetrized counterpart $Q_s=S^{1/2}QS^{-1/2}=-iS^{-1/2}(dS/dE)S^{-1/2}$ introduced in~\cite{brou97a}. The two matrices have the same eigenvalues, i.e. the same proper delay times. The advantage of $Q_s$ is that it has the natural symmetry appropriate to the symmetry class under consideration: $Q_s$ is a real symmetric ($\beta=1$), complex Hermitian ($\beta=2$), or quaternion self-dual matrix ($\beta=4$). This makes $Q_s$ the natural variable for formulating its invariant distribution. Moreover, for ideal coupling $Q_s$ is statistically independent of $S$, so that the joint density factorizes as
\begin{equation}\label{P(S,Q)}
  \mathcal{P}_{\overline{S}=0}(S,Q_s) = P_0(S)P_Q(Q_s)
  \propto   \theta(Q_s)\,\det(Q_s)^{-3\beta M/2-2+\beta}
  e^{-(\beta/2)t_H\,\mathrm{tr}\,Q_s^{-1}}.
\end{equation}
Here $\theta(Q_s)$ denotes the matrix Heaviside function, restricting the support of the distribution to positive-definite matrices $Q_s>0$. By contrast, the non-symmetrized matrix $Q$ retains correlations with $S$ through its eigenvectors in the orthogonal and symplectic cases ($\beta=1,4$).

The Laguerre form~\eqref{jpdf_td} immediately implies strong level repulsion between proper delay times and heavy tails in marginal distributions, with an algebraic decay $\propto\tau^{-\beta M/2-2}$ for large times $\tau\gg t_H$. Such power-law tails are a universal feature of time-delay distributions at finite $M$. Physically, they reflect rare events in which a scattering state remains anomalously long in the interaction region, and can be related to universal statistics of small resonance widths~\cite{fyod97}. The eigenvalue density corresponding to~\eqref{jpdf_td} can be computed by standard orthogonal-polynomial methods, with explicit expressions available for $\beta=2$ at arbitrary $M$ and for $M\gg1$ at arbitrary $\beta$. In the latter case, the limiting form of the distribution of the scaled proper delay time $t=\tau/t_H$ is governed by the Marchenko-Pastur law \cite{brou99}
\begin{equation}\label{dist_proper}
  P_{M\gg1}(t\equiv\tau/t_H) = \frac{1}{2\pi t^2}\sqrt{(t_+-t)(t-t_-)}\,, \qquad t_{\pi}=(3\pm\sqrt{8})/M,
\end{equation}
for $t_-\leq t\leq t_+$, vanishing outside this interval. This result reflects the Wishart--Laguerre structure of the random time-delay matrix, ultimately originating from the factorized representation~\eqref{Q}. On the other hand, the statistics of partial time delays is different. In the case considered, they become statistically equivalent to the diagonal elements of the symmetrized matrix $Q_s$~\cite{savi01}. The corresponding rates (inverse partial time delays) are $\chi^2$ distributed with $M\beta$ degrees of freedom. so that the (scaled) partial time-delay distribution is given by
\begin{equation}\label{dist_partial}
  P_{M}(\tilde{t}\equiv\tilde\tau/t_H) = \frac{(\beta/2)^{\beta M/2}}{M\Gamma(\beta M/2)}
  \tilde{t}^{\,-2-\beta M/2} e^{-\beta/(2\tilde{t})}\,.
\end{equation}
This distribution has the same algebraic tail as the proper-delay distribution at finite $M$, but differs in its full form because partial delays do not exhibit the eigenvalue repulsion characteristic of proper delay times. The distinction is especially clear for large $M$: proper delay times occupy a finite support, Eq.~\eqref{dist_proper}, whereas partial delay times become sharply concentrated around their mean.

The matrix density~\eqref{P(S,Q)} can also be used to derive the distribution of the Wigner time delay. In the $\beta=2$ case, a determinantal representation is available for arbitrary $M$, while explicit formulae for all three symmetry classes are known at $M=1,2$. For $M=1$, all time-delay characteristics coincide, and the distribution is therefore given by Eq.~\eqref{dist_partial}. For $M=2$, one obtains~\cite{savi01}
\begin{equation}\label{dist_wigner}
  P_{M=2}(t_W\equiv\tau_W/t_H) =
  \frac{\beta^{3\beta+2}\Gamma(3(\beta+1)/2)}{\Gamma(\beta+1)\Gamma(3\beta+2)}
  t_W^{-3(\beta+1)} U\left(\frac{\beta+1}{2},2(\beta+1),\frac{\beta}{t_W}\right) e^{-\beta/t_W}\,,
\end{equation}
where $U(a,b,z)$ is Tricomi's (confluent hypergeometric) function. As the number of channels $M$ grows, the Wigner time-delay distribution approaches a Gaussian form in the bulk, with $P_{M\gg1}(t_w)\sim \exp[-\frac{\beta}{8}M^2(Mt_W-1)^2]$ for $t_W\sim M^{-1}$~\cite{mezz13}. This is expected for the linear statistics $\tau_W$ in view of the sum rule \eqref{td_sumrule}. The full limiting distribution, however, exhibits a much richer structure. Using Coulomb gas and large deviation methods applied to the joint density~\eqref{jpdf_td}, one finds the following asymptotic behaviors for small and large times~\cite{texi13}:
\begin{equation}\label{dist_winner_M}
  P_{M\gg1}(t_W) \propto t_W^{-3\beta M^2/4}e^{-\beta M/(2t_W)}\quad(t_W\to0)  \qquad\mbox{and}\qquad
  P_{M\gg1}(t_W)\propto (Mt_W-1)^{-2-\beta M/2}\quad
  (Mt_W-1\gg\sqrt{(2/M)\ln{M}}).
\end{equation}
In analogy with transport problems, the Coulomb gas undergoes a phase transition separating the Gaussian bulk from the far-tail regime.

The Laguerre representation~\eqref{jpdf_td} also provides a starting point for studying moments, cumulants and related time-delay characteristics. All second-order statistics of delay times are known exactly at finite $M$ for all three symmetry classes. In particular, explicit formulae are available for the variances and covariances of the proper delay times~\cite{mezz11,mart14}, for the corresponding quantities for partial delay times~\cite{Kuipers2014}, and for the variance of the Wigner time delay~\cite{mezz13}. Written in normalized form, the relative variances are
\begin{equation}\label{var_tau}
\frac{\mathrm{var}(\tau)}{\aver{\tau}^2} = \frac{M[\beta(M-1)+2]+2}{(M+1)(\beta M-2)}\,,\qquad  \frac{\mathrm{var}(\tilde\tau)}{\aver{\tilde\tau}^2} = \frac{2}{\beta M-2}\,,\qquad
\frac{\mathrm{var}(\tau_W)}{\aver{\tau_W}^2} = \frac{4}{(M+1)(\beta M-2)}\,.
\end{equation}
These expressions display markedly different large-$M$ scalings: the relative variance of a proper delay time remains of order unity, that of a partial delay time scales as $M^{-1}$, while that of the Wigner time delay scales as $M^{-2}$. This reflects the distinct asymptotic behaviour of the corresponding distributions discussed above. Higher cumulants of the Wigner delay time are also accessible; in particular, its skewness and kurtosis are known exactly~\cite{mezz13}. As in the case of transport distributions, these results can be used to construct Edgeworth approximations to the Wigner time-delay distribution, which remain accurate in the bulk even for relatively small numbers of channels.

Several approaches have recently been developed for the systematic study of higher time-delay moments, such as $\mathrm{Tr}\,Q^k$, and their joint statistics. Because of the universal algebraic tail $\sim \tau^{-\beta M/2-2}$ of the time-delay distributions, such moments exist only for $k<\beta M/2+1$. The Selberg-integral and symmetric-function methods used in transport problems have analogues for the Laguerre ensemble and lead to exact combinatorial representations for moments of proper delay times \cite{mezz11,mezz12}. For the Wigner time delay, the integrable-theory approach provides an efficient recursion scheme for the cumulant generating function \cite{mezz13}. In the large-$M$ limit, the joint statistics of quantities of the form $\mathrm{Tr}\,Q^k$ exhibit asymptotic Gaussian behaviour for suitable collections of time-delay linear statistics \cite{Cunden2015prb,Cunden2016}. Finally, averages of polynomial functions of $Q$ can be expressed in terms of Schur polynomials for $\beta=2$ \cite{Novaes2015} and Jack polynomials for general $\beta$ \cite{Novaes2022}.

\subsubsection{Nonideal coupling}

In the general case of nonideal coupling, exact finite-$M$ results for several time-delay characteristics can be obtained by the supersymmetry method~\cite{verb85}. This includes the distribution of partial delay times~\cite{fyod97,savi01}, including the GOE--GUE crossover regime~\cite{fyod97a}, as well as the distribution of proper delay times~\cite{somm01} and its generalization to systems with uniform absorption~\cite{savi03a}. The distribution of the full time-delay matrix, however, is not directly accessible within that approach. To overcome this difficulty, an analogue of the Poisson kernel for the time-delay problem was developed in~\cite{Grabsch2018}. It provides the joint distribution of $S$ and the symmetrized matrix $Q_s$ for arbitrary coupling.

The basic idea is parallel in spirit to the reduction of non-ideal to ideal coupling for the scattering matrix itself. One starts from an ideally coupled cavity, for which $S_0$ is circularly distributed and $Q_{s0}$ has the Laguerre distribution, and then composes this slow chaotic scattering stage with a prompt scattering stage representing the tunnel barriers. The resulting transformation relates the nonideal pair $(S,Q_s)$ to the ideal pair $(S_0,Q_{s0})$,  whose joint distribution is known. A useful feature of this mapping is that the corresponding matrix differentials satisfy $d[S]\,d[Q_s^{-1}]=d[S_0]\,d[Q_{s0}^{-1}]$, which may be viewed as a matrix analogue of Liouville's theorem. In the case of equivalent channels, when the average $S$-matrix is proportional to the identity, $\overline{S}_{ab}=\overline{S}\,\delta_{ab}$, the joint distribution takes the following compact form:
\begin{equation}\label{P(S,Q)kappa}
  \mathcal{P}_{\overline{S}}(S,Q_s)  \propto \theta(Q_s)\,\left|\det(1-\overline{S}^*S)\right|^{\beta M}
   \det(Q_s)^{-3\beta M/2-2+\beta} \exp\left\{-\frac{\beta\,t_H}{2(1-|\overline{S}|^2)}
   \mathrm{tr}\left[(1-\overline{S}^*S)(1-\overline{S}S^\dagger)Q_s^{-1}\right]\right\}.
\end{equation}
For $\overline S=0$, this expression reduces to the perfect-coupling result~\eqref{P(S,Q)}, as expected.

The distribution of the symmetrized time-delay matrix is obtained by integrating Eq.~\eqref{P(S,Q)kappa} over $S$. This integration becomes tractable after parametrizing the scattering matrix as $S=(1-i\kappa K)(1+i\kappa K)^{-1}$, where the coupling strength $\kappa$ is related to $\overline{S}$ via \eqref{T}, and $K$ denotes Wigner's reaction matrix at perfect coupling. The latter has the matrix Cauchy distribution $P(K)\propto\det(1+K^2)^{-1-\beta(M-1)/2}$. One then obtains
\begin{equation}\label{P(Q)kappa}
  P(Q_s)  \propto \theta(Q_s)\,\det(Q_s)^{-3\beta M/2-2+\beta} \int d[K] \frac{\det(1+K^2)}{\det(1+\kappa^2K^2)} \exp\left\{-\frac{\beta}{2} \kappa t_H\mathrm{tr}\left[\frac{1+K^2}{1+\kappa^2K^2}Q_s^{-1}\right]\right\}\,,
\end{equation}
where the integration is over the set of Hermitian matrices with real ($\beta=1$), complex ($\beta=2$) or quaternionic ($\beta=4$) entries.
In the $\beta=2$ case, the JPDF of the eigenvalues of $Q_s$ can be further extracted by integration over the unitary group. Unequally coupled channels can also be treated within the same framework, although the resulting expressions are less compact. Thus Eqs.~\eqref{P(S,Q)kappa} and \eqref{P(Q)kappa} interpolate continuously between the ideal-coupling Laguerre distribution ($T=1$) and the weak-coupling regime when the transmission coefficient $T<1$.

One important application concerns the Wigner time delay. In view of the sum rule~\eqref{td_sumrule}, $\tau_W$ is a linear statistics of the time-delay matrix $Q_s$. This makes it possible to use the matrix distribution~\eqref{P(Q)kappa} to derive and analyse the characteristic function of $\tau_W$. In the weak-coupling limit $MT\ll1$, the distribution of $\tau_W$ develops several distinct asymptotic regimes. Denoting $t_W=\tau_W/t_H$, one finds~\cite{Grabsch2018}
\begin{equation}\label{distr_wigner_weak}
  P(t_W)  \propto \left\{
  \begin{array}{ll}
  t_W^{-(\beta M^2+3)/2} e^{-\beta M T /(8t_W)}, & t_W \ll T \\[1ex]
  T^{1/2} t_W^{-3/2}, & T \ll t_W \ll (M^2T)^{-1} \\[1ex]
  T^2M^3 (TM^2 t_W)^{-2-\beta M/2}, & t_W \gg (M^2T)^{-1}
 \end{array}\right..
\end{equation}
At very small delay times, the distribution has a large-deviation form with an essential exponential suppression. In the intermediate range, it displays a superuniversal $\tau^{-3/2}$ behaviour, independent of both the symmetry class and the number of channels. The same asymptotic law also appears for partial~\cite{fyod97,fyod97a} and proper~\cite{somm01} delay times, making it a robust signature of time-delay statistics in the weak-coupling regime. At larger times, the distribution crosses over to a universal power-law tail whose exponent depends on $\beta$ and $M$. These regimes reflect the interplay between weak escape through the contacts and fluctuations of resonance widths, showing how nonideal coupling enhances the probability of very large delay times through anomalously narrow resonances.

Finally, we note that a related deformation of the ideal-coupling result arises when the system remains perfectly coupled to the observable channels but is subject to uniform absorption. The scattering matrix is then subunitary, and the finite-absorption analogue of the time-delay matrix can be introduced through the unitarity deficit $1-S^\dagger S$, which measures the absorbed flux, as discussed in~\cite{savi03a}. The corresponding distribution of the Wigner--Smith matrix was analysed in \cite{Grabsch2020}, where uniform absorption was implemented by coupling the cavity to a large number of weak fictitious channels while keeping their total transmission fixed. Although this mechanism is physically distinct from nonideal coupling discussed above, it again leads to a deformation of the Laguerre ensemble. In this setting, one can derive a matrix-integral representation for the joint distribution of proper delay times at finite absorption and further describe its large-$M$ properties in terms of coupled Coulomb gases \cite{Grabsch2020}. This provides a complementary controlled extension of the ideal-coupling theory, particularly relevant for microwave and other wave-chaotic systems where losses are unavoidable; see Sec.~\ref{sec:appl} for further discussion.

\section{Correlation properties and resonance statistics}\label{corr}

The maximum-entropy approach provides an efficient description of fixed-energy scattering statistics and, in particular, of transport observables a given energy. Its scope is, however, limited by its single-energy nature. Correlation properties of the $S$-matrix at different energies, as well as spectral characteristics of open systems encoded in the resonances, require information about the energy dependence of $S(E)$ and therefore cannot be obtained from the Poisson-kernel formulation alone. To address such quantities, one has to return to the effective Hamiltonian representation~\eqref{S2}. Combined with the supersymmetry method for ensemble averaging, this formulation has led to exact non-perturbative results for two-point correlation functions of $S$-matrix elements and many other characteristics, as discussed below.

\subsection{Energy correlations of $S$-matrix elements}\label{sec:Scorr}

A central object is the connected energy correlation function of $S$-matrix elements,
\begin{equation}\label{Scorr}
C^{abcd}_S(\omega)\equiv\aver{S^{ab*}_{\mathrm{fl}}(E_1)S^{cd}_{\mathrm{fl}}(E_2)}
=\int_0^{\infty}\!\! d t\, e^{2\pi i\omega t} \hat{C}^{abcd}_S(t)\,,
\end{equation}
where $S_{\mathrm{fl}}=S-\overline{S}$ denotes the fluctuating part of the $S$-matrix. It is natural to  measure the energy separation in units of the mean level spacing $\Delta$, setting $\omega=(E_2-E_1)/\Delta$. In the RMT limit $N\to\infty$, the correlation function depends on the two energies only through this dimensionless frequency. The Fourier transform $\hat{C}^{abcd}_S(t)$ then describes a gradual loss of correlations in time $t$, measured in units of the Heisenberg time $t_H$. Causality implies $\hat{C}^{abcd}_S(t)=0$ for $t<0$. Physically, this time-domain correlator is related to the probability current through a surface enclosing the scattering region and may therefore be interpreted as a decay law of the open system \cite{ditt00}.

The energy dependence of $S(E)$ is made explicit by its pole expansion, $S^{ab}(E)=\delta^{ab}-i\sum_n w_n^a\tilde{w}_n^b/(E-\mathcal{E}_n)$, which follows from Eq.~\eqref{S2}. Here $\mathcal{E}_n$ are the complex resonance poles, while $w_n^a$ and $\tilde w_n^b$ denote the corresponding coupling amplitudes entering the residues. These quantities are defined by the projections of the left and right eigenvectors of $\Heff$ onto the channel vectors. Because of the unitarity constraints imposed on $S$, the pole positions and residues are not statistically independent but develop nontrivial correlations \cite{soko89}. Consequently, the knowledge of the JPDF of resonances alone, even when available, is insufficient for calculating the correlation function~\eqref{Scorr}. The supersymmetry method provides the appropriate framework for carrying out the required ensemble average. In their seminal work~\cite{verb85}, Verbaarschot, Weidenm\"uller and Zirnbauer obtained the exact result for~\eqref{Scorr} at arbitrary transmission coefficients in the orthogonal symmetry class. The corresponding result for unitary symmetry was later derived in \cite{fyod05r}. We summarize these results below.

The exact analytic expression for the correlation function can be written in the universal tensor form
\begin{eqnarray}\label{ScorrFT}
 C^{abcd}_S = \delta^{ab}\delta^{cd}T_aT_c\sqrt{(1{-}T_a)(1{-}T_c)} J_{ac} + (\delta^{ac}\delta^{bd}+\delta_{1\beta}\delta^{ad}\delta^{bc})T_aT_bP_{ab}\,.\ \ \
\end{eqnarray}
The same decomposition applies in the time domain to $\hat{C}^{abcd}_S(t)$, so the argument of the functions $J_{ac}$ and $P_{ab}$ can be suppressed. The term proportional to $\delta_{1\beta}$ accounts for the additional symmetry relation $S^{ab}=S^{ba}$ in the presence of time-reversal invariance. The functions $J_{ac}$ and $P_{ab}$ contain the nontrivial dynamical information and are represented, in the energy domain, as certain expectation values in the field theory (zero-dimensional nonlinear supersymmetric $\sigma$-model). In the orthogonal case ($\beta=1$), one obtains
\begin{equation}\label{SSgoe}
\begin{array}{l}
 \displaystyle
 J_{ac}(\omega) =  \Bigl\langle
 \Bigl(\sum\limits_{i=1}^{2}\frac{\mu_i}{1+T_a\mu_i}+\frac{2\mu_0}{1-T_a\mu_0}\Bigr)
 \Bigl(\sum\limits_{i=1}^{2}\frac{\mu_i}{1+T_c\mu_i}+\frac{2\mu_0}{1-T_c\mu_0}\Bigr)
 \mathcal{F}_M \Bigr\rangle_{\mu}^{\mathrm{goe}} \\[1ex]
 \displaystyle
 P_{ab}(\omega) = \Bigl\langle \Bigl(
 \sum\limits_{i=1}^{2}\frac{\mu_i(1+\mu_i)}{(1+T_a\mu_i)(1+T_b\mu_i)}
 +\frac{2\mu_0(1-\mu_0)}{(1-T_a\mu_0)(1-T_b\mu_0)} \Bigr) \mathcal{F}_M
 \Bigr\rangle_{\mu}^{\mathrm{goe}}
\end{array},
\quad\mbox{with }
\mathcal{F}^{}_{M}=\prod_c\frac{1-T_c\mu_0}{\sqrt{(1+T_c\mu_1)(1+T_c\mu_2)}}\,.
\end{equation}
In the unitary case ($\beta=2$), the corresponding expressions read
\begin{equation}\label{SSgue}
\begin{array}{l}
 \displaystyle
 J_{ac}(\omega) = \Bigl\langle
 \Bigl(\frac{\mu_1}{1+T_a\mu_1}+\frac{\mu_0}{1-T_a\mu_0}\Bigr)
 \Bigl(\frac{\mu_1}{1+T_c\mu_1}+\frac{\mu_0}{1-T_c\mu_0}\Bigr)
 \mathcal{F}_M\Bigr\rangle_{\mu}^{\mathrm{gue}} \\[1ex]
 \displaystyle
 P_{ab}(\omega) = \Bigl\langle \Bigl( \frac{\mu_1(1+\mu_1)}{(1+T_a\mu_1)(1+T_b\mu_1)}
 +\frac{\mu_0(1-\mu_0)}{(1-T_a\mu_0)(1-T_b\mu_0)} \Bigr)\mathcal{F}_M\Bigr\rangle_{\mu}^{\mathrm{gue}}
\end{array} ,
\quad\mbox{with } \mathcal{F}_{M}=\prod_c\frac{1-T_c\mu_0}{1+T_c\mu_1}\,.
\end{equation}
The averages $\langle(\cdots)\rangle_{\mu}$ introduced above are to be understood explicitly as
\begin{equation}
 \langle(\cdots)\rangle_{\mu} = \left\{
 \begin{array}{ll}
  \displaystyle
   \frac{1}{8} \int_{0}^{\infty}\!d\mu_1 \!\int_{0}^{\infty}\!d\mu_2 \!\int_{0}^{1}
   \frac{d\mu_0 \,(1-\mu_0)\mu_0 |\mu_1-\mu_2|\,\exp[i\pi\omega(\mu_1+\mu_2+2\mu_0)] }{ [(1+\mu_1)\mu_1(1+\mu_2)\mu_2]^{1/2} (\mu_0+\mu_1)^2 (\mu_0+\mu_2)^2 }
   (\ldots)\,, & \mbox{for GOE}  \\[2ex]
 \displaystyle
   \int_{0}^{\infty}\!\!d\mu_1\int_{0}^{1}\!d\mu_0
   \frac{ \exp\{i2\pi\omega(\mu_1+\mu_0)\} }{ (\mu_1+\mu_0)^{2} }(\ldots)\, & \mbox{for GUE}
 \end{array}
 \right. .
\end{equation}
In these integral representations, the dimensionless time is identified with $t=\mu_0+\frac{1}{2}(\mu_1+\mu_2)$ for $\beta=1$ and with $t=\mu_0+\mu_1$ for $\beta=2$. The so-called channel factor $\mathcal{F}_M$ carries the essential information about openness, controlling the decay of long-time scattering correlations.

Several important limiting cases allow the general expressions above to be simplified. We first mention the case of elastic scattering with only one open channel. In this regime the correlation function has a strongly non-Lorentzian profile. More generally, for a finite number $M$ of open channels the long-time behaviour of the form factor is algebraic, $\hat C^{abcd}_S(t)\sim t^{-M\beta/2-2}$. Such power-law decay is a characteristic feature of open chaotic systems with a finite number of decay channels \cite{ditt00}. Formally, it follows from the large-$t$ asymptotics of the channel factor, $\mathcal{F}_M\sim\prod_c(1+\frac{2}{\beta}T_ct)^{-\beta/2}$. Physically, this behaviour reflects fluctuations of resonance widths, which are especially pronounced for few open channels and become progressively weaker as the number of channels grows; see further discussion in Sec.~\ref{sec:poles} below.

A qualitatively different limit is obtained when the number of open channels is large, $M\gg1$, while each channel is weakly transmitting, $T_c\ll1$, with $\sum_c T_c$ kept finite. In this semiclassical regime the individual width fluctuations are self-averaged, and resonances acquire essentially the same escape width $\gamma_T=\sum_{c=1}^{M}T_c$ (in units of $t_H^{-1}$), the so-called Weisskopf width. Correspondingly, the channel factor becomes exponential, $\mathcal{F}_M\simeq\exp[-\gamma_Tt]$. The remaining fluctuations of the scattering matrix are then governed by the spectral correlations of the corresponding closed system, which enter through the canonical two-level RMT form factor $b_{2,\beta}(t)$, resulting in
\begin{equation}\label{ScorrMT}
 C^{abcd}_S(\omega) =
 \frac{(\delta^{ac}\delta^{bd}+\delta_{1\beta}\delta^{ad}\delta^{bc})T_aT_b
 }{ \gamma_T-2\pi i\omega } + \delta^{ab}\delta^{cd}T_aT_c
 \int_{0}^{\infty}\!\!dt\, [1-b_{2,\beta}(t)]  e^{-(\gamma_T-2\pi i\omega)t}\,,
 \qquad\gamma_T = \sum_c^{M}T_c\,.
\end{equation}
In the Ericson regime of strongly overlapping resonances, $\gamma_T\gg1$, the first term gives the dominant Lorentzian contribution, whereas the second term represents the correction due to spectral correlations of the closed system.

As a direct application, one obtains predictions for the experimentally accessible elastic enhancement factor,
$\mathrm{var}(S^{aa})/\mathrm{var}(S^{ab})$, $a\neq b$, from the $S$-matrix correlation function at $\omega=0$. In particular, the dominant Lorentzian term in Eq.~\eqref{ScorrMT} yields the well-known Hauser--Feshbach relation in the Ericson regime. The above results have been also generalized to the full crossover of gradually broken time-reversal invariance (the GOE--GUE crossover), and the resulting predictions have been tested experimentally in chaotic microwave billiards \cite{diet09}.

The same formalism can also be applied to \emph{half-scattering} (or \emph{half-collision}) processes, i.e. to decaying quantum systems prepared by an external excitation rather than by a scattering event. Important examples include photodissociation and atomic autoionization, where photon absorption excites the system into an energy range with chaotic dynamics, followed by decay into open channels. Using the optical theorem, autocorrelation functions of photodissociation cross-sections can be related to correlations of $S$-matrix elements. Of particular interest are Fano resonances, which arise from the interference between long-time chaotic decay and short-time direct escape. This interference can again be described in terms of the $S$-matrix correlations discussed above; see \cite{gori05} for detailed discussion and further references.

Finally, we mention recent RMT studies of scattering in two coupled chaotic subsystems separated by a barrier, each coupled to its own set of external channels. Such models provide a useful framework for describing transport between two chaotic regions mediated by a transition state. This is closely related to transition-state theory in physics and chemistry, where one seeks to compute reaction and decay rates for complex many-particle systems in the presence of a barrier. For further background and recent developments, see \cite{hagino2024microscopic,weidenmuller2024transition}.

\subsection{Time-delay correlations}\label{sec:td}

The average Wigner time delay is fixed by the mean density of states,
$\langle\tau_W\rangle=2\pi/(M\Delta)=t_H/M$. Its fluctuations around this value are described by the autocorrelation function of the fluctuating part, $\tau_W^{\mathrm{fl}}=\tau_W-\langle\tau_W\rangle$. Exact RMT expressions were obtained for the orthogonal case in \cite{lehm95b} and for the unitary case in \cite{fyod97}. In the notation used in Eqs.~\eqref{SSgoe}--\eqref{SSgue}, they read as follows:
\begin{equation}\label{td_corr}
 C_Q(\omega) \equiv \frac{ \langle\tau_W^{\mathrm{fl}}(E-\frac{\Delta}{2}\omega) \tau_W^{\mathrm{fl}}(E+\frac{\Delta}{2}\omega)\rangle }{ \langle\tau_W\rangle^2} = \left\{ \begin{array}{l}
  2\re\left\langle (2\mu_0+\mu_1+\mu_2)^2\mathcal{F}_M\right\rangle_\mu^{\mathrm{goe}} \\[1ex]
  \re\left\langle (\mu_0+\mu_1)^2\mathcal{F}_M \right\rangle_\mu^{\mathrm{gue}}
 \end{array} . \right.
\end{equation}
The corresponding result is also known in the full GOE--GUE crossover, including the generalization to parametric correlations~\cite{fyod97a}.

Representation~\eqref{td_corr} has the same structural origin as the $S$-matrix correlations discussed above and can be analysed in a similar way. In particular, the time-delay correlator in the semiclassical many-channel regime reduces to
\begin{equation}\label{td_corrMT}
C_Q(\omega) = 2\mathrm{Re}\int_0^\infty dt\,[1-b_{2,\beta}(t)]e^{-(\gamma_T-2\pi i\omega)t}
 \approx\frac{4}{\beta\gamma_T^2}
 \frac{1-(2\pi\omega/\gamma_T)^2 }{[1 + (2\pi\omega/\gamma_T)^2]^2}\,,
 \qquad\mbox{for } \gamma_T\gg1.
\end{equation}
Thus, in contrast to the leading Lorentzian profile of the $S$-matrix correlations in the Ericson regime, the time-delay correlator has a derivative-Lorentzian form. The depletion and sign change encoded in~\eqref{td_corrMT} are manifestations of the correlation hole, originating from the interplay between exponential decay in the open system and spectral rigidity in its closed counterpart.

The same formula also connects naturally to the variance of the Wigner time delay. Setting $\omega=0$ in~\eqref{td_corrMT} gives
\begin{equation}\label{td_var}
  \frac{\mathrm{var}(\tau_W)}{\langle\tau_W\rangle^2} \equiv C_Q(0)
  \simeq \left\{
  \begin{array}{ll}
    2/\gamma_T \gg 1\,,  &  \mbox{for } \gamma_T \ll 1 \\[1ex]
    4/(\beta\gamma_T^2) \ll 1\,, & \mbox{for } \gamma_T\gg 1
  \end{array} \right. .
\end{equation}
Thus the relative variance gets parametrically suppressed when crossing from isolated to overlapping resonances. These limits are consistent with the behaviour inferred from the Wigner time-delay distribution: weak coupling produces broad fluctuations dominated by narrow resonances, whereas in the Ericson regime the fluctuations are strongly reduced by self-averaging over many decay channels.

\subsection{$S$-matrix poles and residues}\label{sec:poles}

The correlation functions discussed above encode the energy dependence of scattering observables on the real axis. A complementary, and in many respects more intrinsic, description is obtained by analytically continuing $S(E)$ into the complex plane. Within the effective-Hamiltonian formalism, the poles of $S(E)$ are the complex eigenvalues $\mathcal{E}_n=E_n-\frac{i}{2}\Gamma_n$ of $\Heff$, while the residues are determined by the corresponding left and right eigenvectors. These quantities   provide a direct spectral characterization of the open chaotic system. In particular, resonance position and widths determine decay rates and time-delay fluctuations, whereas the residues encode the nonorthogonality of resonant states and their sensitivity to perturbations. We now review the main RMT results for these pole and residue statistics.

\subsubsection{Resonance statistics in elastic chaotic scattering}
To analyse systematically the statistical properties of the $N$ complex eigenvalues $\{\mathcal{E}_n\}$ of $\Heff$, it is natural to start with their JPDF, denoted by $\mathcal{P}_M(\mathcal{E})$. Unfortunately, such a density is known in full generality only in rather special cases. The most important one is elastic scattering, $M=1$, when the anti-Hermitian part of $\Heff$ has rank one. Denoting its only nonzero eigenvalue by $-i\kappa$, the effective Hamiltonian becomes a rank-one non-Hermitian deformation of the Hermitian matrix $H$. For this problem the JPDF was already derived by Ullah in 1969 \cite{ulla69}, and was later rediscovered and further analysed in the context of chaotic scattering, notably in \cite{soko89}. A unified derivation valid for arbitrary Dyson index $\beta>0$ was given more recently in \cite{kozhan2017rank}. Assuming $\kappa>0$ to be non-random, one finds in the GOE case
\begin{equation} \label{w1a}
 \mathcal{P}_{M=1}^{\mathrm{goe}}(\{\mathcal{E}_i\}) \propto \prod_{k=1}^N
 \frac{e^{-\frac{N}{4}|\mathcal{E}_k|^2} }{ \sqrt{\im\mathcal{E}_k}} \frac{|\Delta(\{\mathcal{E}_i\})|^2}{\prod\limits_{m<n}|\mathcal{E}_m{-}\mathcal{E}^*_n|} \frac{e^{-\frac{N}{4}\kappa^2}}{\kappa^{N/2-1}} \, \delta\biggl(\kappa+\sum_{i=1}^N \im\mathcal{E}_i\biggr)\,,
\end{equation}
whereas in the GUE case the JPDF is given by
\begin{equation} \label{w2a}
 \mathcal{P}_{M=1}^{\mathrm{gue}}(\{\mathcal{E}_i\}) \propto \prod_{k=1}^N e^{-\frac{N}{2}\re\mathcal{E}_k^2}
 |\Delta(\{\mathcal{E}_i\})|^2 \frac{e^{-\frac{N}{2}\kappa^2}}{\kappa^{N-1}}\,\delta\biggl(\kappa+\sum_{i=1}^N \im\mathcal{E}_i\biggr)\,.
\end{equation}
Actually, exploiting a version of the Itzykson-Zuber-Harish-Chandra integral allows one to write the analogue of (\ref{w2a}) at arbitrary fixed number of channels $M$ \cite{fyod99,fyod03r} but the final expression is rather cumbersome.

As is typical for RMT problems, the main challenge is to extract the $n$-point correlation functions associated with the joint densities~\eqref{w1a} and~\eqref{w2a}. The simplest, yet nontrivial statistical characteristics of the resonances is the one-point function, i.e. the mean resonance density. Physically, it can be used to determine the distribution of resonance widths $\Gamma_n$ in a spectral window centred around a given energy $E$. The relevant local energy scale is the mean spacing $\Delta$ between neighbouring resonances along the real axis. One can show \cite{fyod_osm2022} that Eq.~\eqref{w1a} implies the following probability density for the dimensionless widths $y_n=\pi\Gamma_n/\Delta$:
\begin{equation} \label{w3a}
 \rho_{M=1}^{\mathrm{goe}}(y) =
 \frac{1}{4\sqrt{2}e^{-gy}}\,\hat{L}\,\int_1^{\infty}
 e^{-gvy}\frac{v-1}{\sqrt{v+1}}I_0\left(y\sqrt{(g^2-1)(v^2-1)}\right)\,dv\,,
 \qquad\mbox{with } g \equiv \frac{2}{T}-1 =
 \frac{1}{2}\left(\kappa+\frac{1}{\kappa}\right)\,,
\end{equation}
Here $I_0$ is the modified Bessel function, and the differential operator $\hat{L}$ is defined by
$\hat{L}=2\sinh 2y-\left(\cosh 2y-\frac{\sinh 2y}{2y}\right)\left(\frac{3}{y}+2\frac{d}{dy}\right)$.
The coupling parameter $g\geq1$ is related to the transmission coefficient, and hence to $\kappa$,  through Eq.~\eqref{T}. A notable feature of this distribution is the emergence of an algebraic large-width tail $\rho_{M=1}^{\mathrm{goe}}(y)\propto y^{-2}$ in the case of perfect coupling $g=1$.

\subsubsection{Width distribution for finite $M$}
Although the full JPDF of resonances is not known in closed form for general $M$, one can nevertheless derive the marginal distribution of the resonance widths for any finite number of open channels, arbitrary coupling constants $g_c$, and all three symmetry classes $\beta=1,2,4$. The corresponding expressions for general couplings are rather cumbersome (e.g., see Eq.~(3) in \cite{somm99} for the GOE case). In the weak-coupling regime ($T_c\ll1$ or, equivalently, $g_c\gg1$), resonances are narrow and become well separated, with typical widths $\Gamma_n/\Delta\sim T_c\ll1$. In this limit the width statistics reduce to the familiar Porter--Thomas distributions, derived long ago by first-order perturbation theory. Systematic improvement of that approximation was developed more recently in \cite{fyo2015res_pert}.

In the case of equivalent channels, when all $g_c=g$, the resonance width distribution simplifies considerably for arbitrary coupling strength $g\ge1$. For the underlying Hermitian Hamiltonian drawn from the GUE, the corresponding expression is given by~\cite{fyod97}
\begin{equation} \label{den_pol_GUE}
 \rho_{M}^{\mathrm{gue}}(y) = {\cal F}^{(1)}_M(y){\cal F}^{(2)}_M(y), \qquad \mbox{with }
 {\cal F}^{(1)}_M(y)=\frac{y^{M-1}}{(M{-}1)!}e^{-yg} \,\mbox{ and }\,
 {\cal F}^{(2)}_M(y)=(-1)^M e^{yg}\frac{d^M}{dy^M}\left(e^{-yg}\frac{\sinh y}{y}\right)\,.
\end{equation}
This expression has been confirmed by numerical simulations \cite{kott00} and by experiments on chaotic wave scattering on graphs with broken TRI \cite{chen2021stat_poles}. For GOE symmetry of the Hamiltonian, the distribution of resonance widths is also known for arbitrary coupling, but it takes its simplest explicit form in the case of perfect coupling, $g=1$. For arbitrary $M$, one then obtains \cite{Kur25}
\begin{equation} \label{den_pol_GOE}
 \rho_{M}^{\mathrm{goe}}(y) = \frac{1}{2(M{-}1)!}
 \left[1+\frac{d}{dy}\frac{e^y\Gamma\left(\frac{M}{2},y\right)}{2y^{M/2-1}}\right]
 \frac{\gamma\left(M+1,2y\right)}{y^{2}}.
\end{equation}
Here $\gamma(a,y)$ and $\Gamma(a,y)$ denote the lower and upper incomplete gamma functions, respectively. For the GSE case, explicit formulae for $\rho_M^{\mathrm{gse}}(y)$ at perfect coupling were given in \cite{Kur25} for $M=2$ and $M=4$, together with a generating function that can in principle be used to obtain expressions for other even values of $M$.

On can show that, for all symmetry classes and arbitrary coupling strength, the first moment of the distribution (i.e. the mean resonance width) satisfies the Moldauer--Simonius relation
\begin{equation}\label{MS}
\left\langle y\right\rangle = \frac{M}{2}\ln{\left(\frac{g+1}{g-1}\right)}
 = -\frac{M}{2}\ln(1-T)\,.
\end{equation}
This long-known identity, closely related to $S$-matrix analyticity and spectral ergodicity, is very general; see \cite{Kur25} for a recent discussion and references to the original works. Its logarithmic divergence as $T\to1$ ($g\to1$) points out to formation of an algebraic tail $\rho(y)\propto y^{-2}$ in the width distribution at perfect coupling. This behaviour has been observed experimentally in microwave cavities and microwave graphs \cite{kuhl08,chen2021stat_poles}, and is one of the most robust signatures of chaotic resonance statistics.

The emergence of such a power-law tail is closely related to the remarkable phenomenon of \emph{resonance trapping}, predicted in \cite{soko89}, see also \cite{dittes1991formation}, and observed experimentally in chaotic wave scattering \cite{pers00}. To elucidate its origin, recall from Eq.~\eqref{T} that the transmission coefficients $T_c$ are invariant under the transformation $\kappa_c\to\kappa_c^{-1}$. For equivalent channels, with all $\kappa_c=\kappa$, this invariance may suggest a symmetry of the width distribution when $\kappa\to1/\kappa$, which maps weak coupling, $0\leq\kappa\leq1$, to strong coupling, $1\leq\kappa<\infty$. However, a more detailed analysis (see, e.g., the discussion around Eq.~(112) in \cite{fyod97}) reveals that a qualitative reorganization of the $S$-matrix poles takes place when $\kappa$ crosses the perfect-coupling point $\kappa=1$. For fixed $M<N$ and $\kappa<1$, all $N\gg1$ statistically distributed resonances have the widths of order $\Delta$ and follow the finite-$M$ distributions discussed above. For $\kappa>1$, by contrast, $M$ broad resonances separate from the rest and acquire non-random widths of the same order $\Gamma/\Delta\sim\kappa-\kappa^{-1}$, which is large for $\kappa\gg1$. The remaining $N-M$ resonances stay narrow, with the corresponding widths governed by the dual weak-coupling scale $\kappa^{-1}$. This separation into a small number of broad, short-lived states and a large number of trapped narrow, long-lived states is the essence of resonance trapping. For non-equivalent channels the $M$-fold degeneracy of the broad pole is lifted, but the qualitative picture remains unchanged. The resulting restructuring of the pole configuration may be viewed as a collective phenomenon akin to a phase transition. Recently, the formation of such ``gigantic'' resonance widths has been studied rigorously in the simplest rank-one case, $M=1$, identifying the critical scaling regime $|\kappa-1|\sim N^{-1/3}$ of the transition \cite{dubach2023dynamics,fyod_khor_popl_2022extreme}.

\subsubsection{Higher-order correlators (GUE case)}
For systems with broken TRI, more detailed results are available. In the GUE case, and for a fixed channel number $M\ll N$ in the limit $N\to\infty$, one can determine not only the width distribution but all $n$-point correlation functions of the resonance poles in the complex plane. Restricting, for simplicity, to a spectral window centred at $\mathrm{Re}\,\mathcal{E}=0$, these correlation functions acquire the familiar determinantal form
\begin{equation} \label{w6a}
\lim_{N\to \infty}\frac{1}{N^{2n}}R_n\left(z_1=N{\cal E}_1,\ldots,z_n=N{\cal E}_n\right)=\det{\left[K(z_i,z^*_j)\right]^n_{j,k=1}}\,,
\end{equation}
where the kernel is given by \cite{fyod99}
\begin{equation} \label{w6a}
 K(z_1,z^*_2)=F_M^{1/2}(z_1)F_M^{1/2}(z^*_2)\int_{-1}^1 d\lambda\, e^{-i\lambda(z_1-z_2^*)}\prod_{k=1}^M(g_k+\lambda)\,,
\end{equation}
with $F_M(z)=\sum_{c=1}^M \frac{\exp(-2g_c|\im(z)|)}{\prod_{a\ne c}(g_c-g_a)}$. In the particular case of equivalent channels the diagonal $K(z,z^*)$ reproduces the mean density \eqref{den_pol_GOE}.

\subsubsection{Cloud formation and universal regimes for $M\to\infty$}
The preceding results concern the regime of a fixed number of open channels. A different situation arises for chaotic cavities with macroscopic openings, or holes, which in the semiclassical limit support a large number of propagating modes. At the RMT level this motivates considering the limit $M,N\to\infty$ with $m=M/N$ fixed. In this regime the resonance poles form a dense cloud, and sometimes two clouds, in the lower half of the complex energy plane, characterized by a mean density $\rho(z)$ \cite{haak92,lehm95a}. The cloud closest to the real axis is separated from it by a finite gap, whose width sets an important scale controlling scattering correlations \cite{lehm95b,lehm95a}. Quantitatively, however, the detailed shape of the cloud, the gap, and the density profile depend on the choice of the coupling matrix $V$ and are therefore not universal. To isolate universal predictions one instead considers the dilute-opening regime $m=M/N\ll1$. In this limit the density of unscaled resonance poles, $z=X-iY$ with $Y>0$, acquires the following universal form, valid for all three symmetry classes:
\begin{equation}\label{cloud}
\rho(X,Y)=\frac{m}{4\pi Y^2}, \quad \frac{m}{\frac{1}{2}(\kappa+\kappa^{-1})+\pi\nu(X)}\le Y\le \frac{m}{\frac{1}{2}(\kappa+\kappa^{-1})-\pi\nu(X)}, \quad \nu(X)=\frac{1}{\pi}\sqrt{1-\frac{X^2}{4}}
\end{equation}
and $\rho(X,Y)=0$ otherwise. In this leading order approximation we thus reproduce a single sharp-edged resonance cloud  separated by a finite gap $Y_\mathrm{min}=m/[\frac{1}{2}(\kappa+\kappa^{-1})+1]$ from the real axis.

Although the density profile~\eqref{cloud} is ``superuniversal'', in the sense that it is valid for all three standard symmetry classes of the underlying Hamiltonian (GOE, GUE, and GSE), a refined analysis of  the edge behaviour of the resonance cloud reveals universal features that do depend on symmetry. These appear when the limits are taken in the proper order: first $N\to\infty$ at fixed $M<\infty$, and only then $M\gg1$. In this regime, the sharp edge predicted by~\eqref{cloud}, located at $y_{\min}=\frac{M}{2(g+1)}$, is replaced by a smooth crossover envelope. At $X=0$, the edge profile can be written in the rescaled width coordinate as $\rho(y)=\frac{M}{2y^2}f_\beta\left(\frac{y-y_{\min}}{\sqrt{M}}\right)$,
where the scaling function $f_\beta$ depends on the symmetry class. For GUE symmetry, $f_\beta$ follows by extracting the leading large-$M$ contribution from an integral representation of~\eqref{den_pol_GUE}, giving $f_{\beta=2}(z)=\frac{1}{2}\mathrm{erfc}(-\sqrt{2}z)$ in terms of the complementary error function. For GOE and GSE symmetry, the corresponding edge profiles are described by~\cite{Kur25}
\begin{equation}\label{edgeprofile}
  f_{\beta=1}(z)=\left[1+\frac{\sqrt{\pi}}{2}\frac{d}{dz}e^{z^2}
  \mathrm{erfc}{(z)}\right]  f_{\beta=2}(z),
  \qquad
  f_{\beta=4}(z)=\int_{-z}^{\infty}\frac{dx}{\sqrt{8\pi}}e^{-\frac{1}{2}(x-z)^2}
  \left[1+\sqrt{\pi}ze^{x^2}\mathrm{erfc}{(x)}\right].
\end{equation}

The profiles $f_{\beta=2}(z)$ and $f_{\beta=1}(z)$ have appeared previously as edge densities for two major bulk universality classes of strongly non-Hermitian random matrices: class $\mathbf{A}$, corresponding to general complex matrices, and class $\mathbf{AI}^{\dagger}$, corresponding to complex symmetric matrices; see \cite{akemann2025spectral} for discussion and further references. On this basis, it was conjectured in \cite{Kur25} that $f_{\beta=4}(z)$ represents the universal edge profile for the remaining class $\mathbf{AII}^{\dagger}$, corresponding to complex quaternion matrices.

These three symmetry classes are also expected to characterize three types of bulk eigenvalue statistics deep inside the resonance cloud. For an underlying GUE Hamiltonian of the closed system, one can further analyse the $M\to\infty$ limit of the resonance kernel $K(z_1,z_2^*)$ in Eq.~\eqref{w6a}. After appropriate rescaling, the local statistics of resonances inside the cloud are described by a Ginibre-like kernel,
\begin{equation}
 \textstyle
 |K(z_1,z^*_2)|=\rho(z)\exp\{-\frac{1}{2}\pi\rho(z)|z_1-z_2|^2\},\quad z=\frac{1}{2}(z_1+z_2)\,.
\end{equation}
This result is expected to be universal for systems in class $\mathbf{A}$ \cite{fyod03r}. No comparable results are currently available for classes $\mathbf{AI}^{\dagger}$ and $\mathbf{AII}^{\dagger}$.

Finally let us mention that for a superconducting quantum chaotic "billiard dot"  weakly coupled to a normal metal, one comes to the problem of studying resonances in the systems with Andreev reflection \cite{mi2014Beenakker}. Mentioning here for brevity only the simplest case of such systems, at the level of effective Hamiltonian approach it requires to replace the matrix $H$ characterizing the closed dot by a matrix $H=iA$, with real antisymmetric random $A=-A^T$. The complex position of arising resonances possess in this case an additional symmetry: both ${\cal E}_n$ and  $-{\cal E}_n^*$ appear simultaneously as eigenvalues of the effective Hamiltonian. The most interesting consequence of such system is an accumulation of resonances in the imaginary axis.  In the case of many open channels and perfect coupling the corresponding density was found in  \cite{mi2014Beenakker} by mapping on the known eigenvalue density of truncations of orthogonal matrices \cite{khoruzhenko2010truncations}.

\subsubsection{Nonorthogonal resonant states}
As mentioned above, the resonance states, or quasimodes, are identified with the eigenvectors of the non-Hermitian Hamiltonian $\Heff$. Unlike eigenvectors of Hermitian operators, they do not form an orthogonal basis. Rather, the right and left eigenvectors form a biorthogonal system. More precisely, to each complex resonance energy $\mathcal{E}_k$ correspond right and left eigenvectors,
\begin{equation}\label{eig}
 \Heff|R_k\rangle = \mathcal{E}_k|R_k\rangle \quad \mathrm{and} \quad
 \langle L_k|\Heff =\mathcal{E}_k\langle L_k|
\end{equation}
satisfying the conditions of bi-orthogonality, $\langle{L_k|R_m}\rangle=\delta_{km}$, and completeness, $1=\sum_k|R_k\rangle\langle L_k|$. The non-orthogonality manifests itself via the matrix $\mathcal{O}_{mn}=\langle{L_m|L_n}\rangle\langle{R_n|R_m}\rangle$. In the context of reaction theory such a matrix is known as the Bell-Steinberger non-orthogonality matrix which, e.g., influences branching ratios of nuclear cross-sections \cite{soko89} and also features in the particle escape from the scattering region \cite{savi97} and  in the sensitivity of resonance widths to perturbations of scattering system  \cite{fyod_sav2012res_nonorth}.
In the context of quantum optics $\mathcal{O}_{nn}$ yields the enhancement (Petermann's or excess-noise factor) of the line width of a lasing mode in open resonators, whereas $\mathcal{O}_{n\ne m}$  describe cross-correlations between noise emitted in different eigenmodes.

This attracted an essential interest in statistical properties of $\mathcal{O}_{nm}$ in scattering RMT framework.
The mean value ${\cal O}({\cal E})=\langle\sum_n{\cal O}_{nn}\delta({\cal E}_n-{\cal E})\rangle$  of the diagonal nonorthogonality factor is known at $\beta=2$ for any coupling $g$ and any number of channels. Namely, using $y$ for the properly scaled position in the complex plane around which the resonance density is given by $\rho^{gue}_M(y)$ from \eqref{den_pol_GUE}, one finds that \cite{schom00a}:
\begin{equation}\label{mean_overlap}
 {\cal O}(y)=\rho^{gue}_M(y)-\int_0^y{\cal F}_M^{(1)}(u)\frac{d}{du}{\cal F}^{(2)}_M(u)\,du,
\end{equation}
where ${\cal F}^{(1)}_M(y)$ and ${\cal F}^{(2)}_M(y)$ have been defined in \eqref{den_pol_GUE}. For a single open channel the whole distribution of $\mathcal{O}_{nn}$
has been derived in \cite{fyod_osm2022} for both $\beta=1,2$, and shown to be
heavy-tailed, so that only its mean (given for $\beta=2$ by \eqref{mean_overlap} with $M=1$) is finite,
while all moments starting from the variance diverge (see further references in\cite{fyod_osm2022} for related RMT results in Ginibre ensembles) . Further, in the $\beta=2$ single-channel case the off-diagonal correlator ${\cal O}({\cal E},{\cal E}')=
\langle\sum_{n\ne m}{\cal O}_{nm}\delta({\cal E}_n-{\cal E})\delta({\cal E}_m-{\cal E'})\rangle$ is also known non-perturbatively, see \cite{fyod02}.

The off-diagonal elements $\mathcal{O}_{n\neq m}$ contain complimentary information about nonorthogonality. They can be directly linked to the shifts $\delta\Gamma_n$ of resonance widths induced by perturbing the closed counterpart of the open system \cite{fyod12b}. Such shifts provide a sensitive indicator of nonorthogonality, vanishing only if the wavefunctions are orthogonal. Moreover, in weakly open systems ($\Gamma\ll\Delta$), the off-diagonal overlaps are parametrically larger than the self-overlaps: one has $O_{nm}\sim (\Gamma/\Delta)$ \cite{fyod12b}, whereas  $O_{nn}-1\sim (\Gamma/\Delta)^2$ \cite{poli09b}, making $\delta\Gamma_n$ especially sensitive to nonorthogonality effects. The resulting distribution of width shifts represents a distinct type of nonorthogonality statistics; it was derived for both GOE and GUE cases in \cite{fyod12b} and subsequently verified experimentally in weakly open chaotic cavities \cite{gros14}.

One of the other prominent manifestations of eigenfunction nonorthogonality is its influence on decay laws in open wave chaotic systems. A convenient way to quantify this effect is to consider the simplest yet non-trivial characteristics, the decay of the total field intensity inside the scattering domain after an initial excitation, often referred to as \emph{norm leakage}. It can be represented as \cite{savi97}:
\begin{equation}\label{P(t)}
 P(t) \equiv \frac{1}{N} \Bigl\langle \mathrm{Tr} \Bigl( e^{i\Heff^{\dag}t}e^{-i\Heff t}\Bigr) \Bigr\rangle
  = \frac{1}{N} \Bigl\langle \sum_{n,m} \mathcal{O}_{mn} e^{i(\mathcal{E}^*_n-\mathcal{E}_m)t} \Bigr\rangle\,.
\end{equation}
The first representation makes clear that $P(t)$ can be viewed as a natural open-system analogue of the Loschmidt echo: the forward evolution with $\Heff$ is followed by the backward evolution with $\Heff^\dagger$, and the two do not cancel because $\Heff$ is non-Hermitian. For a closed (Hermitian) system $P(t)=1$ identically at all times. In an open system, however, the anti-Hermitian part of $\Heff$, which describes coupling to the continuum, produces a mismatch between the two evolutions and hence a nontrivial decay. The second representation shows that this decay is governed jointly by the finite resonance widths and by the nonorthogonality of the resonant states.

The exact analytic expression for $P(t)$ obtained by supersymmetry calculations \cite{savi97,savi03a} is given by the function $P_{ab}(t)$ entering the $S$-matrix correlation functions, see Eqs.~\eqref{SSgoe} and~\eqref{SSgue}, where one has to set the channel transmissions $T_a$ and $T_b$ appearing explicitly in the denominators to zero. Its typical behaviour is given by the channel factor, $P(t)\sim\prod_c(1+\frac{2}{\beta}T_c t)^{-\beta/2}$. In the so-called \emph{diagonal approximation}, which neglects the nonorthogonality of resonance states and becomes asymptotically exact at large $t$, $P(t)$ is related directly to the width distribution by the Laplace transform $P_{\mathrm{diag}}(t)=\frac{1}{N}\langle\sum_n e^{-\Gamma_n t}\rangle=\int_{0}^{\infty}d\Gamma\,e^{-\Gamma t}\rho(\Gamma)$. A detailed analysis of the time evolution and the associated time scales in decaying chaotic systems shows that resonance state nonorthogonality mainly affects the early stages of the decay \cite{savi97}. Qualitatively, this is because the large-time behaviour is governed by narrow resonances with small widths, which are effectively isolated and for which nonorthogonality enters only perturbatively (see \cite{poli09b} for the related study). Relations between $S$-matrix pole statistics and decay or relaxation laws have been further discussed in various regimes and models of quantum chaos; see, e.g., \cite{casa97,Kolovsky_chapter}.

\section{Other characteristics and applications}\label{sec:appl}

We have so far focused mainly on universal statistics of scattering characteristics in flux-conserving systems. In many experimental realizations, however, losses are unavoidable or even central to the measurement protocol. Microwave cavities, acoustic and elastic billiards, optical resonators and complex electromagnetic enclosures all involve absorption. In mesoscopic conductors one may also encounter dephasing and finite-frequency response. In such situations, part of the incoming flux is irreversibly lost in the environment, resulting in the observable $S$-matrix that is no longer unitary. The RMT approach based on the effective Hamiltonian is naturally suited to incorporate such losses. In this section we discuss this generalization and related applications to several measurable observables.

\subsection{Incorporating uniform absorption}\label{sec:abs}

Physically, absorption corresponds to dissipative decay of the wave intensity inside the scattering region. Although different spectral components may in general have different dissipation rates, this weak energy dependence can be neglected when one is interested in local fluctuations on the much finer scale $\sim\Delta$. In this  approximation of uniform absorption, all resonances acquire the same additional absorption width $\Gamma_\mathrm{abs}>0$, on the top of their escape widths $\Gamma_n$. The corresponding dimensionless absorption strength is $\gamma \equiv t_H\Gamma_\mathrm{abs} = 2\pi\Gamma_\mathrm{abs}/\Delta$, with $\gamma\ll1$ or $\gamma\gg1$ describing weak or strong absorption, respectively. Microscopically, absorption can be modelled either by coupling the system to a complicated background with almost continuous spectrum or, equivalently, by introducing a large number of weakly open fictitious channels and taking the limit $M_{\mathrm{abs}}\to\infty$, $T_{\mathrm{abs}}\to0$, with $M_{\mathrm{abs}}T_{\mathrm{abs}}=\gamma$ kept fixed; see \cite{fyod05r,savi03a} for discussion.

Operationally, uniform absorption is equivalent to a purely imaginary shift of the scattering energy, $E \to E+\frac{i}{2}\Gamma_\mathrm{abs}\equiv E_\gamma$. The observable scattering matrix is then $S_\gamma(E)\equiv S(E_\gamma)$ and is subunitary. The missing flux is conveniently quantified by the unitarity deficit $1-S_\gamma^\dagger S_\gamma$, which can be represented at arbitrary $\Gamma_\mathrm{abs}$ as follows \cite{savi03a}:
\begin{equation}\label{SdagS}
  1-S_\gamma^\dagger(E) S_\gamma(E) = \Gamma_\mathrm{abs}Q_\gamma(E)\,,\qquad
  Q_\gamma(E) \equiv V^{\dagger}\frac{1}{(E_\gamma-\Heff)^{\dagger}}\frac{1}{E_\gamma-\Heff} V\,.
\end{equation}
This representation provides a natural finite-absorption generalization of the time-delay matrix $Q_\gamma$, as a measure of the amount of incoming flux absorbed inside the system. Taking the trace and performing the statistical averaging, one can further obtain the exact relation \cite{savi03a}
\begin{equation}\label{SdagS_trace}
  1-\frac{1}{M} \aver{\mathrm{Tr}(S_\gamma^\dagger S_\gamma)}
  = \frac{\Gamma_\mathrm{abs}}{M}\aver{\mathrm{Tr}Q_\gamma}
  = \frac{\gamma}{M}\left(1-\gamma\int_{0}^{\infty}dt\,e^{-\gamma t} P(t)\right)\,,
\end{equation}
which expresses the average unitarity deficit in terms of the norm-leakage decay function~\eqref{P(t)}. It shows explicitly how absorption reduces the effective dwell time due to dissipation in the environment. In the case of vanishing absorption, $Q_\gamma$ reduces to the conventional Wigner--Smith time-delay matrix. Relation~\eqref{SdagS} therefore provides an experimental route to time-delay statistics, extending the earlier proposal \cite{doro90} for extracting the time-delay distribution in a single-channel weakly absorbing cavity. This connection was used to derive a number of exact RMT results for time-delay statistics, reflection and related probabilities of no return in chaotic cavities with absorption, see \cite{fyod05r} for the overview and further discussion. It also shows that absorption need not be regarded merely as a nuisance: it can serve as a diagnostic resource by converting otherwise inaccessible dynamical information into measurable reflection statistics. In particular, absorption leads to a complex extension of the Wigner time delay, which can be exploited to infer resonance statistics from scattering measurements \cite{chen2021stat_poles}.

For correlation functions, uniform absorption enters in an especially simple way. The imaginary energy shift produces an additional exponential damping factor in the Fourier time representation. Thus the time-domain form factors discussed above acquire a factor $e^{-\gamma t}$, which suppresses the universal algebraic tails $\sim t^{-2-M\beta/2}$ for large times $t\ge\gamma^{-1}$ (in units of $t_H$). This provides a practical way to incorporate homogeneous losses into scattering correlations and is widely used in the analysis of microwave-cavity data, where the absorption strength can be extracted from the decay of measured correlation functions \cite{schaf03}. In particular, extending the results of Sec.~\ref{sec:Scorr} to finite absorption gives analytic expressions for elastic enhancement factors \cite{fyod05r,savi06a}, which have been verified experimentally \cite{zhen06,lawn10,Bialous2023}.

More substantial changes occur at the level of distributions, where absorption alters the nature of the basic observables. The scattering matrix becomes subunitary, while the Wigner reaction matrix $K_\gamma\equiv K(E_\gamma)$ evaluated at the complex energy is no longer Hermitian. In statistical electromagnetics, the matrix $iK_\gamma$ is interpreted as the normalized cavity impedance \cite{grad14,hemm05}, so RMT predictions translate directly into impedance statistics \cite{fyod04b}. A general fluctuation-dissipation type relation further connects these finite-absorption quantities to correlation properties of $K$ in the corresponding lossless system \cite{savi05}. This relation provides a basis for a powerful nonperturbative approach \cite{fyod05r} to study impedance and reflection coefficient statistics in chaotic cavities, which has been successfully used to interprete microwave and network experiments with absorption \cite{kuhl05,Hemmady2012,lawn19}. We briefly discuss this developement and recent extensions next.

\subsection{$K$-matrix (impedance) statistics}\label{sec:K}

The $K$ matrix depends on channel couplings multiplicatively, so universal results are most naturally formulated at perfect coupling (all $\kappa_c=1$ and setting $E=0$). In the absence of absorption, $K$ is Hermitian and follows the matrix Cauchy (Lorentzian) ensemble, cf. Eq.~\eqref{P(Q)kappa}. At finite absorption, $K_\gamma$ becomes non-Hermitian and its full matrix distribution is not known in closed form. Nevertheless,  exact results are available for several marginal statistics, including diagonal \cite{fyod05r} and off-diagonal entries \cite{fede20} as well as the complex eigenvalues \cite{fyodorov2023APPA}.

The diagonal elements are of particularly importance for applications. From Eq.~\eqref{S1}, a diagonal entry $(K_\gamma)_{nn}=\frac{N\Delta}{\pi}\langle n|(E_\gamma-H)^{-1}|n\rangle$
is given the rescaled local Green function of the closed system at the point of channel attachment. By rotational invariance of the RMT ensemble, this point may be chosen arbitrarily. It is therefore sufficient to consider the single-channel variable $K_\gamma=u-iv$, where $v>0$ is the local density of states, normalised as $\aver{v}=1$. Importantly, $u$ and $v$ are correlated random variables with the following JPDF \cite{fyod04b}:
\begin{equation}\label{P(u,v)}
 \mathcal{P}(u,v)=\frac{1}{2\pi v^2}P_0(x), \qquad x=\frac{u^2+v^2+1}{2v}> 1.
\end{equation}
This form is invariant under $iK\to(iK)^{-1}$, thus implying that the normalized impedance and admittance have the same distribution. The function $P_0(x)$ has the meaning of a probability distribution and can be related to that of the reflection coefficient, $r=|S_\gamma|^2$. Namely, using the polar form for $S_\gamma=(1-iK_\gamma)/(1+iK_\gamma)=\sqrt{r}e^{i\theta}$, one readily finds $r=\frac{x-1}{x+1}$. At perfect coupling, the scattering phase $\theta$ is uniformly distributed and independent of $r$, leading to the universal form~\eqref{P(u,v)}, which is therefore valid for all RMT symmetry classes.

The explicit evaluation of $P_0(x)$ can be carried out using a general relation between the joint distribution~\eqref{P(u,v)} at finite absorption and an analytically continued energy correlator of resolvents of the lossless $K$ matrix \cite{savi05}. In one convenient form, this relation reads
\begin{equation}\label{P(u,v)int}
  \int_{-\infty}^{\infty}du \int_{0}^{\infty}dv \frac{\mathcal{P}(u,v)}{(z'-u)^2+(z''-v)^2}
  = \aver{
  \frac{1}{z_{-}-K(-\Omega/2-i0)}\frac{1}{z_{+}-K(\Omega/2+i0)}}_{\Omega=i\Gamma_\mathrm{abs}
  },  \qquad z_+ = (z_-)^* = z'+iz'' \ \ (z''>0).
\end{equation}
This identity has the character of a fluctuation--dissipation relation: the left-hand side describes the dissipative distribution of the complex $K_\gamma$ matrix at finite absorption, whereas the right-hand side is expressed through a response-type correlator of the corresponding lossless system. The latter can be evaluated by the supersymmetry method, similarly to the calculation of $S$-matrix correlations discussed in Sec.~\ref{sec:Scorr}. This yields an exact analytical result that is valid throughout the GOE--GUE crossover of gradually broken TRI \cite{savi05}. In the GOE case the final expressions simplify only partially and remain rather involved. In the GUE case, however, one obtains the compact explicit result $P_0^{\mathrm{gue}}(x)=\frac{1}{2}\left[\frac{\gamma}{2}(x+1)(e^\gamma-1)+1+\gamma-e^\gamma\right]e^{-\gamma(x+1)/2}$,
which can also be derived by an alternative method \cite{fyod04b}.

Once the joint density~\eqref{P(u,v)} is known, one can obtain the marginal distributions of the real and imaginary parts of $K_\gamma$ by integration. In statistical electromagnetics, $u$ and $v$ are interpreted as the normalized cavity reactance and resistance, respectively \cite{grad14}. This provides a direct route for comparing the theory \cite{fyod05r} with microwave experiments; see \cite{hemm05,Hemmady2012,lawn19} for further details.

A similar strategy can be applied to the complex eigenvalues $k_1,\ldots,k_M$ of the non-Hermitian matrix $K_\gamma$ at finite absorption for arbitrary $M$. Writing $k_c=u_c-iv_c$, their mean eigenvalue density retains the same functional structure as~\eqref{P(u,v)} for all RMT symmetry classes \cite{fyodorov2023APPA}:
\begin{equation}
 \mathcal{P}_M(u,v) \equiv \left\langle \frac{1}{M}\sum_{c=1}^M \delta(u-u_c)\delta(v-v_c) \right\rangle
 = \frac{1}{2\pi v^2}P_M(x), \quad x=\frac{u^2+v^2+1}{2v}.
\end{equation}
This structure reflects the statistical independence of the moduli and phases of the complex eigenvalues $s_c=\sqrt{\frac{x_c-1}{x_c+1}}\,e^{i\theta_c}$, $c=1,\ldots,M$, of the matrix $S_\gamma$, with phases uniformly distributed. In the GUE case, the distribution $P_M(x)=\left\langle \frac{1}{M}\sum_{c=1}^M\delta(x-x_c)\right\rangle$
is known explicitly:
\begin{equation}
  P_M^{\mathrm{gue}}(x) = \frac{1}{2}\frac{d}{d x} \left[(x-1)\phi_M\left(\frac{\gamma}{2}(x+1)\right)-(x+1)\phi_M\left(\frac{\gamma}{2}(x-1)\right)\right],
  \qquad \phi_M(a)=e^{-a} \frac{1}{M}\sum_{k=0}^{M-1}\frac{(M-k)!}{k!}a^k
\end{equation}
For $M=1$, this expression reduces to the single-channel function $P_0(x)$ given above. Extensions to other symmetry classes, as well as connections to the statistics of $S$-matrix poles, are under current investigation and will be published elsewhere.

Finally, we mention that the distribution of off-diagonal entries of the $K$ matrix is also known analytically \cite{fede20}. In applications, these entries describe cross-port impedance fluctuations and can therefore be probed experimentally in multiport microwave networks and related wave chaotic systems; see \cite{lawniczak2020experimental,farooq2024coupled} for examples and further discussion.

\subsection{Distribution of $S$-matrix entries}\label{sec:distr_S}

Another class of experimentally accessible quantities is provided by the individual entries of the scattering matrix. In particular, the diagonal entries describe reflection amplitudes, while the off-diagonal entries describe transmission between distinct channels. In the ideal, lossless case these entries are strongly constrained by unitarity. In the presence of absorption, however, the scattering matrix becomes subunitary and the moduli of its entries acquire nontrivial distributions that depend on the absorption strength and on the coupling to the channels. The analytical treatment of diagonal and off-diagonal entries is quite different. Diagonal elements are directly related to the reaction matrix and can often be studied through the distribution of $K$. Off-diagonal elements, by contrast, involve correlations between different channels and generally require more elaborate computational techniques. We therefore discuss the two cases separately.

\subsubsection{Diagonal elements}\label{sec:distr_S_diag}

For a single channel, the $S$ and $K$ matrices reduce to scalar quantities. They are related by $S=(1-iK)/(1+iK)$ in the absence of absorption, with the continuation $K\to K_\gamma$ at finite absorption. Thus the distribution of the scattering amplitude can be obtained directly from the distribution $\mathcal{P}(u,v)$ of $K_\gamma=u-iv$ discussed above. For nonideal coupling, $T\leq1$, direct reflection at the channel opening interferes with the wave reflected from the chaotic interior. As a result, the scattering phase is no longer uniformly distributed and becomes statistically correlated with the reflection coefficient $r$ (or $x$). Nevertheless, their joint density is still determined by the same function $P_0$ \cite{fyod04b}:
\begin{equation}\label{P(x,theta)}
  P(x,\theta) = \frac{1}{2\pi}P_0\left(xg - \sqrt{(x^2-1)(g^2-1)}\cos\theta\right)\,,\qquad g = 2/T-1\geq1.
\end{equation}
This representation provides an efficient way to analyse reflection statistics in absorbing microwave cavities \cite{kuhl05}.

For several channels, the diagonal element $S_{aa}$ can be isolated by separating channel $a$ from the remaining open channels,
\begin{equation}\label{S_aa}
  S_{aa}(E) = \frac{1-iK_a(E)}{1+iK_a(E)}\,,
  \qquad K_a(E) = \frac{1}{2} V_a^\dagger\frac{1}{E-\Heff^{(a)}}V_a\,,
\end{equation}
where $\Heff^{(a)}=H-\frac{i}{2}\sum_{c\neq a}V_c V_c^\dagger$ is now non-Hermitian. Thus the statistics of the diagonal scattering amplitude, and hence of the reflection coefficient in channel $a$, is reduced to the joint distribution of $K_a=u_a-iv_a$, with $v_a>0$.

This distribution can be obtained by the same methods as in Sec.~\ref{sec:K}. Technically, the effect of the other channels enters through an additional `channel factor', which accounts for their arbitrary coupling strengths, while the selected channel $a$ is treated as perfectly coupled. Nonideal coupling in channel $a$ can then be restored by the transformation~\eqref{P(x,theta)}. In this way one obtains the general joint distribution of the reflection amplitude and phase. The resulting framework also yields explicit distributions in a number of physically relevant regimes, including cases where localization effects begin to modify the universal chaotic-scattering statistics; see \cite{fyod05r} for details and discussion.

\subsubsection{Off-diagonal elements}\label{sec:distr_S_off}

The distribution of off-diagonal elements, $S_{ab}$ with $a\neq b$, is more challenging. Unlike diagonal reflection amplitudes, transmission amplitudes between distinct channels cannot be reduced to a scalar reaction-matrix variable. They involve correlations between different channel wave functions inside the scattering region and are therefore sensitive to the full matrix structure of the effective Hamiltonian approach.

The supersymmetry method provides a systematic way to compute such distributions \cite{nock14}. The resulting expressions are typically given as integral representations rather than elementary functions. For GUE systems with broken TRI the distribution of the modulus of an off-diagonal element can be written in terms of a small number of auxiliary integration variables and channel-dependent coupling constants. The GOE case is technically more involved, reflecting the additional correlations imposed by the symmetry $S_{ab}=S_{ba}$. Although the final formulae are more cumbersome, they can be evaluated numerically and compared directly with experimental data.

We only consider the simpler GUE case here. Let $r=|S_{ab}|$ denote the modulus of an off-diagonal matrix element. Without loss of generality, we may take $a=1$ and $b=M$. Introducing the coupling parameters $g_c=2/T_c-1\geq1$, the probability density $P_r(r)$ in this symmetry class, normalized as $\int_0^\infty P_r(r)\,r\,dr=1$, can be represented as \cite{nock14}
\begin{equation}\label{Nock1}
 P_r(r)=\frac{1}{r}\frac{\partial}{\partial r} r\frac{\partial}{\partial r} f(r), \quad
  f(r)=\frac{1}{2}\frac{(g_1+\lambda_1)^2(g_M+\lambda_1)^2}{(g_1+g_M)\lambda_1^2+2(g_1g_M+1)\lambda_1+(g_1+g_M)}\, {\cal U}(r)
\end{equation}
and the function ${\cal U}(r)$ is given by
\begin{equation}
 \label{Nock2a}
 {\cal U}(r)=\int_{-1}^1\frac{d\lambda_2}{(\lambda_1-\lambda_2)^2}\prod_{c=1}^M\frac{g_c+\lambda_2}{g_c+\lambda_1} \left(\frac{\lambda_1^2-1}{(g_1+\lambda_1)(g_M+\lambda_1)}+
 \frac{1-\lambda_2^2}{(g_1+\lambda_2)(g_M+\lambda_2)}\right),
\end{equation}
with $\lambda_1$ for a given $r$ being defined via
\begin{equation}\label{Nock3}
 \lambda_1=\frac{(g_1+g_M)r^2+\sqrt{(g_1-g_M)^2r^4+4r^2(g_1g_M-1)+4}}{2(1-r^2)}.
\end{equation}
This representation is well suited for analysing the dependence of transmission-amplitude fluctuations on channel coupling and absorption.

Physically, off-diagonal distributions interpolate between different regimes. In weakly open or weakly absorbing systems, transmission is dominated by isolated resonances, leading to broad, strongly non-Gaussian distributions. In strongly absorbing or many-channel regimes, many resonant contributions overlap strongly (Ericson regime), and the complex transmission amplitude approaches a Gaussian random variable, with Rayleigh-type statistics for its modulus. The exact RMT expressions describe the crossover between these regimes and therefore provide a quantitative benchmark for wave-chaotic scattering experiments. We note that integral representations for the distribution of $|S_{ab}|$ in systems with preserved TRI were also derived in \cite{nock14}. They are more involved and will not be reproduced here. Nevertheless, they can be evaluated numerically and have been compared successfully with microwave-billiard experiments \cite{kuma13}, including more recent exact results and measurements in the Ericson-fluctuation regime \cite{kohnes2025exact,kohnes2026universality}.

\subsubsection{Fluctuations in an established transmission}\label{sec:distr_S_est}

In many wave transport settings, such as wireless communication or microwave links, one is interested in a transmission channel that is already established by design. The role of the surrounding complex environment is then not to create the transmission, but to induce fluctuations around an otherwise deterministic signal. A useful non-perturbative description of this situation combines a simple transmitting mode coupled to a chaotic background \cite{savi17}. At the resonance energy of the mode, the resulting scattering matrix can be written as
\begin{equation}\label{S_est}
  S = 1 - \frac{1}{1+i\eta K_\gamma}(1-S^{(0)}), \qquad  \eta \equiv \Gamma_{\downarrow}/\Gamma_0\,,
\end{equation}
where $S^{(0)}$ describes the deterministic (or direct) transmission process, $\Gamma_0$ is the natural escape width of the transmitting mode, and $\Gamma_{\downarrow}$ is the so-called spreading width describing its coupling to the chaotic background. The dimensionless parameter $\eta$ therefore measures the strength of environmental mixing relative to direct escape. The chaotic background acts effectively as a random scattering center, dephasing the deterministic transmission and inducing fluctuations in both transmission and reflection. A crucial distinction from the cases discussed earlier is that the average scattering matrix is now generically non-diagonal, $\langle S\rangle=\frac{1}{1+\eta}(\eta+S^{(0)})$. Consequently, the exact results for distributions of individual $S$-matrix entries reviewed above, which assume a diagonal $\aver{S}$, cannot be applied directly.

Representation~\eqref{S_est} nevertheless makes the problem tractable in full generality. It reduces the statistical analysis to that of the (scalar) reaction matrix $K$, whose statistics at finite absorption have been discussed above. In this way, the non-diagonal structure of the optical scattering matrix is absorbed into the deterministic matrix $S^{(0)}$, while the random part is controlled by the universal distribution of $K$. In particular, for a lossless system at perfect coupling ($T=1$) one obtains the distribution of the transmission intensity $t$ as
\begin{equation}\label{P_est_t}
  P_{\gamma=0}(t)=
  \frac{1}{\pi\sqrt{t(1-t)}\,[\eta t+\eta^{-1}(1-t)]}\,.
\end{equation}
This expression displays a characteristic bimodal structure, reflecting the competition between direct transmission and environmental mixing. Finite absorption smooths this shape and produces exponential cutoffs near both edges, $t\to0$ and $t\to1$. The exact distribution $P_\gamma(t)$ at arbitrary absorption can be expressed through the function $P_0(x)$ entering the joint density~\eqref{P(u,v)} above; see \cite{savi17} for details. A particularly striking consequence is that the established transmission is strongly suppressed when the spreading width exceeds the escape width, $\eta>1$. This regime marks the dominance of dephasing by the chaotic background and loss of information about the original transmission signal.

This approach can be further extended by considering not only the transmission intensity but the full complex transmission amplitude. In \cite{savi18}, the joint distribution of the transmission envelope and phase was derived for arbitrary coupling to the chaotic background and finite absorption. The result shows that the intensity and phase are strongly correlated within a finite domain of support. These correlations remain important even at strong absorption, where one might otherwise expect a simple Gaussian or Rician description to apply. In that limit, a compact asymptotic expression was obtained which gives a uniformly accurate approximation throughout the full nontrivial support of the distribution, thereby improving on standard phenomenological approximations.

In the similar way, one can also derive the joint distribution of reflection and transmission in this problem \cite{savi20}. Notably, the distribution exhibits a remarkable symmetry between its reflection and transmission sectors at perfect coupling, controlled by the same parameter $\eta$. This symmetry persists at the level of the marginal densities even though flux conservation is broken by absorption. The approach also allows one to study the loss statistics in the scattering process. For example, the distribution of the total loss $d=1-r-t$ can be written as
\begin{equation}\label{Ploss}
  P_{\eta}(d) =
  \int_{0}^{\pi}\frac{d\theta}{\pi d^2}
  P_0\left[
  \frac{1}{d}\left(
  g_\eta(1-d)
  - \sqrt{(g_\eta^2-1)(1-2d)}\cos\theta
  \right)\right],
  \qquad
  g_\eta \equiv \frac{\eta+\eta^{-1}}{2}\geq1 .
\end{equation}
In the GOE case this representation is already the most useful form. In the GUE case, the remaining angular integration can be performed explicitly, yielding a closed expression for the loss distribution at arbitrary absorption and arbitrary coupling to the chaotic background. These results provide a quantitative theory of how a deterministic transmission mode is degraded by a complex lossy environment.

We also mention a complementary way of incorporating complex wave environments, recently proposed in the context of targeted mode transport \cite{Wang2025}. Instead of following the fluctuations of a prescribed transmission mode coupled to a chaotic background, it focuses on the optimal energy transfer between selected input and output channels in a multimode cavity. This viewpoint is particularly relevant for wavefront shaping and communication applications, where the complex environment is exploited as a resource for controlled transport.

\subsection{Distribution of local field intensity} \label{distr_int}

The incoming channel amplitudes generate a field inside the scattering region. Within the present framework, this field is represented by a vector $\boldsymbol{\psi}$ in the $N$-dimensional internal Hilbert space, given by the linear superposition $\boldsymbol{\psi}=\sum_{c=1}^M a_c\,\Psi_c$ of the internal parts introduced in Eq.~\eqref{Q}. In a cavity setting, it is convenient to use the position basis $|{\bf r}\rangle$ associated with a suitable coordinate system inside the cavity, so that $\psi({\bf r})\equiv\langle {\bf r}|\boldsymbol{\psi}\rangle$ is the local wave amplitude at point ${\bf r}$ and $\mathcal{I}_{\bf r}=|\psi({\bf r})|^2$ is the corresponding field intensity. Summing the intensities over all internal basis states and using the completeness relation $\sum_{\bf r}|{\bf r}\rangle\langle{\bf r}|=\hat 1_N$, one finds
$\sum_{\bf r}\mathcal{I}_{\bf r}={\bf a}^\dagger Q(E)\,{\bf a}$,
where $Q(E)$ is the Wigner--Smith time-delay matrix. In this sense, $Q(E)$ measures the total internal intensity generated by a given incoming wave.

The representation in terms of $\Heff$ also provides a natural starting point for studying the probability density $P_M(\mathcal{I})$ of the local intensity $\mathcal{I}=|\psi({\bf r})|^2$ at a fixed observation point inside the scattering region. To isolate universal behaviour, the observation point is assumed to lie sufficiently far from the channel openings, so that non-universal near-contact effects can be neglected. For systems with broken time-reversal invariance ($\beta=2$), the corresponding distribution was derived in \cite{fyodorov2023intensity} and admits the following convenient parametrization. For a given $\mathcal{I}>0$, let $\lambda_1>1$ be the unique solution of
\begin{equation}\label{Igeneq}
\mathcal{I}=\frac{\lambda_1-1}{2}\sum_{c=1}^M |a_c|^2\left(1-\frac{g_c-1}{\lambda_1+g_c}\right)\,.
\end{equation}
Introducing further the notation $\tilde g_c=\frac{1+g_c\lambda_1}{g_c+\lambda_1}$, the probability density can then be written in terms of $\lambda_1$ as
\begin{equation}\label{mainresult_general}
P_M(\mathcal{I}) = \frac{d}{d\mathcal{I}}\mathcal{I}\frac{d}{d\mathcal{I}}
\sum_{c=1}^M|a_c|^2 \mathcal{F}_c(\mathcal{I})\,, \quad
\mathcal{F}_c(\mathcal{I}) = \frac{ \lambda_1-1 }{ \biggl(2\mathcal{I}+(\lambda_1-1)^2\sum\limits_{i=1}^M|a_{i}|^2\frac{g_{i}-1}{(\lambda_1+g_{i})^2}\biggr)\prod\limits_{j = 1}^M(\lambda_1+g_{j})}
  \int_{-1}^{1}\,d\lambda_2\,
\, \frac{\lambda_2+\tilde{g}_c}{\lambda_1-\lambda_2}\prod_{k\ne c }^M(\lambda_2+g_{k})\,.
\end{equation}

In the special case when the incoming waves are fed to the system via only a single channel, one can show that the distribution of intensities $P_M(\mathcal{I})$ can be actually recovered from the distribution of $S-$matrix elements \eqref{Nock1} via a certain limiting procedure \cite{fyodorov2023intensity}, but in a general case the two statistics
are independent. The expression for  $P_M(\mathcal{I})$ above takes the most simple form if all channels are perfectly coupled with $g_c=1$. In that case after defining the total incoming flux  as
$I_0=\sum_{c=1}^M|a_c|^2$ one finds
\begin{equation}\label{perfect1}
  P_M(\mathcal{I})=(M+1)\frac{I_0^{M+1}}{(I+I_0)^{M+2}}.
\end{equation}
In fact the tail behaviour can be easily shown to have the same powerlaw  form $P_M(\mathcal{I})\sim \mathcal{I}^{-(M+2)}$ for any coupling. We thus conclude that in the absence of internal absorption for any finite number of open channels $M<\infty$ the ensuing powerlaw-tailed distribution is quite different from the  Rayleigh law  for the intensity distribution predicted by the so-called ``Gaussian random wave'' model \cite{brouwer2003wave}.  Note however that setting in Eq.(\ref{perfect1}) the number of channels to infinity in such a way that the incoming flux per channel remains finite: $\lim_{M\to \infty}{\cal I}/M=\overline{I}<\infty$ restores the Rayleigh law: $\lim_{M\to \infty}P_M(\mathcal{I})=\frac{1}{\overline{\cal I}}e^{-\mathcal{I}/\overline{\cal I}}.$ This fact supports the view that the Gaussian wave model is asymptotically accurate if scattering system is  open in an essentially semiclassical way, with many incoming channels supporting finite flux per channel.  One can further find the joint probability density of intensities $I_1, \ldots, I_L$  in several observation points inside the scattering domain, and then extract the corresponding statistics for the maximal intensity in the observation pattern.  For $L\to \infty$ the resulting limiting  extreme value statistics (EVS) turns out to be different from the classical EVS distributions \cite{fyodorov2023intensity}.

\subsection{Quantum maps and sub-unitary random matrices}

The scattering approach can be easily adopted to treat open dynamical systems with discrete time, i.e. open counterparts of the so-called area-preserving chaotic maps.  The later are usually represented by unitary operators which act on Hilbert spaces of finite large dimension $N$, being often referred to as evolution, scattering or Floquet operators, depending on the given physical context. Their eigenvalues (eigenphases) consist of $N$ points on the unit circle and conform statistically quite accurately the results obtained for Dyson's circular ensembles. Making these systems "open" by allowing escape from "holes" in the phase space provides a useful framework for studying resonance-like phenomena in such "leaky maps", see \cite{glue02,jacq03,schom09} for diverse physical applications,   \cite{nova13a} for a review.

 A general scattering approach framework for such systems was developed in \cite{fyod00} and we mention its gross features below.
For a closed linear system characterized by a wavefunction $\Psi$  the ``stroboscopic'' dynamics amounts to a linear unitary map such that $\Psi(n+1)=\hat{u}\Psi(n)$. The unitary evolution operator $\hat{u}$ describes the inner state domain decoupled both from input and output spaces. Then a coupling that makes the system open must convert the evolution operator $\hat{u}$ to a contractive operator $\hat{A}$ such that $1-\hat{A}^{\dagger}\hat{A}\ge 0$. The equation $\Psi(n+1)=\hat{A}\Psi(n)$ describes now an irreversible decay of any initial state $\Psi(0)\ne 0$ when an input signal is absent. On the other hand, assuming a nonzero input and zero initial state $\Psi(0)=0$, one can relate the (discrete) Fourier-transforms of the input and output signals at a frequency $\omega$ to each other by a $ M\times M$  unitary scattering  matrix $\hat{S}(\omega)$ as follows:
\begin{equation} \label{S-discr}
 \hat{S}(\omega)=\sqrt{1-\hat{\tau}^{\dagger}\hat{\tau}}-\hat{\tau}^{\dagger}
 \frac{1}{e^{-i\omega}-\hat{A}}\hat{u}\hat{\tau}\,, \qquad \hat{A}=\hat{u} \sqrt{1-\hat{\tau}\hat{\tau}^{\dagger}}\,,
\end{equation}
where $\hat{\tau}$ is a rectangular $N\times M$ matrix with $M\le N$ nonzero entries $\tau_{ij}=\delta_{ij}\tau_j$, $0\le \tau_i\le 1$. This formula is a complete discrete-time analogue of Eq.~(\ref{S2}). In particular, one can straightforwardly verify unitarity and show that
\begin{equation}\label{dets2}
 \det{\hat{S}(\omega)}=e^{-i\omega N}
 \frac{\det{\left(\hat{A}^{\dagger}-e^{i\omega}\right)}}
 {\det{\left(e^{-i\omega}-\hat{A}\right)}}
 = e^{-i\omega N}\prod_{k=1}^N
 \frac{\left(z_k^{*}-e^{i\omega}\right)}{\left(e^{-i\omega}-z_k\right)},
\end{equation}
where $z_k$  stand for the complex eigenvalues of the matrix $\hat{A}$ (note $|z_k|<1$). This relation is an obvious analogue of Eq.(\ref{dets1}) and gives another indication of $z_k$ playing the role of resonances for the discrete time systems.

Generic features of quantized maps with chaotic inner dynamics are emulated by choosing $\hat{u}$ from one of the Dyson circular ensembles. By averaging Eq.~(\ref{S-discr}) over $\hat{u}$, one easily finds $\hat{\tau}^{\dagger}\hat{\tau}=1-|\langle\hat{S}\rangle |^2 $. Therefore, $M$ eigenvalues $T_a=\tau_a^2\le1$ of $\hat{\tau}^{\dagger}\hat{\tau}$ play the familiar role of transmission coefficients. In the particular case of all $T_{a}=1$ (ideal coupling), the non-vanishing eigenvalues of the matrix
$\hat{A}$ coincide with those of a $(N{-}M){\times}(N{-}M)$ subblock of $\hat{u}$. Complex eigenvalues of such ``truncations'' of random unitary matrices were first studied analytically in \cite{zycz00}, with corresponding RMT results playing a very useful benchmark role for comparison to spectra of "leaky maps", see recent work in this direction in   \cite{signor2025beyond}. The statistics of complex eigenvalues for a more general ensemble of $N{\times}N$ random contractions $\hat{A}=\hat{u}\sqrt{1-\hat{\tau}\hat{\tau}^{\dagger}}$,  with $u\in CUE$, can be studied starting from the following equivalent probability measure in the matrix space ($d\hat{A}=\prod d{\re A}_{ij}d{\im A}_{ij}$):
\begin{equation}\label{0}
 \mathcal{P}(\hat{A})d\hat{A} \propto \delta(\hat{A}^{\dagger}\hat{A}-\hat{G}) d\hat{A}\,,\quad
 \hat{G}\equiv {\bf 1} -\hat{\tau}\hat{\tau}^{\dagger}\,,
\end{equation}
see \cite{fyod03r} for more detail and ensuing spectral characteristics.

\subsection{Microwave cavities with localized losses}

We have discussed above the approximation of uniform absorption, which is often adequate for descrining homogeneous Ohmic losses in microwave billiards with imperfectly conducting walls. In many experimental settings, however, losses may be spatially localized, e.g., in complex reverberant structures, room-temperature microwave cavities, or in cavities with deliberately introduced absorbers. Unlike uniform absorption, which shifts all resonances by the same imaginary amount, localized losses produce mode-dependent broadenings determined by the spatial overlap of each mode with the lossy region. Within the effective Hamiltonian framework, such losses can be incorporated naturally by treating them as additional absorption channels. In contrast to the fictitious-channel model of uniform absorption, where a large number of weak channels is taken with fixed total transmission, localized losses correspond to a finite number of absorptive channels with finite coupling strength. This distinction is important, as the resulting loss operator has finite rank and hence probes the spatial structure of the internal wave functions. Such inhomogeneous damping affects the statistics of transmitted power in multichannel dissipative ergodic structures \cite{rozh03}, and it also leads to complexness (non-orthogonality) of cavity modes \cite{savi06b}. More recently, the shape of reflected signals from lossy chaotic cavities has been proposed as a probe of eigenfunction non-orthogonality \cite{fyod_osm2022}.

Localized losses are not merely  experimental imperfection but can also be used as a resource. This viewpoint is central to the design of coherent perfect absorption (CPA), where a lossy cavity acts as a perfect interference trap for suitably prepared incident coherent radiation \cite{chong2010coherent}. In this sense CPA is often regarded as the time-reversed counterpart of lasing. In complex enclosures, the condition for CPA is highly sensitive to the interplay between external coupling, chaotic internal dynamics, and the spatial profile of losses. Since localized absorption is the key ingredient, the effective Hamiltonian formulation with finite-rank loss operators provides a natural RMT framework to address CPA-related effects. Such an approach was initiated perturbatively in \cite{li2017random}, and later developed into a full non-perturbative treatment in \cite{fyodorov2017CPA,osman2020chaotic}, linking CPA to zeros of the subunitary scattering matrix and extending the resonance pole viewpoint to absorptive systems.

With localized losses modelled by the absorptive channels, as discussed above, the measured scattering matrix is obtained by restricting the full unitary scattering matrix to the observable channels. In other words, localized losses are represented by unobserved channels, while the experimentally accessible matrix $S_A(E)$ is the corresponding observable subblock. In the Hamiltonian approach this amounts to supplementing the physical channel coupling matrix $V$ by an absorptive coupling matrix $V_A$. Defining the positive semi-definite loss operator $\Gamma_A=V_A V_A^\dagger$, the observable subunitary $S$-matrix admits two equivalent representations, generalizing Eqs.~\eqref{S1} and~\eqref{S2}:
\begin{equation}\label{S_ABS}
 S_A(E) = \frac{1-iK_A(E)}{1+iK_A(E)} = 1-iV^{\dag} \frac{1}{E+i\epsilon-\Heff} V \,,
 \qquad\mbox{with }\, \Heff = H - \textstyle\frac{i}{2}(VV^{\dag}+\Gamma_A)
 \,\mbox{ and }\, K_A(E) = \textstyle\frac{1}{2}V^{\dag} \left(E+\frac{i}{2}\epsilon-H+\frac{i}{2}\Gamma_A\right)^{-1} V \,.
\end{equation}
Here $\epsilon\geq0$ accounts for possible additional spatially uniform absorption, while $\Gamma_A$ describes localized losses. The mismatch between incoming and outgoing fluxes is now quantified by the unitary deficit matrix as follows
\begin{equation}\label{Udef}
 1- S_A(E)^{\dagger}S_A(E) = V^{\dagger}\frac{1}{E-i\epsilon-\Heff^{\dag}}(\epsilon+\Gamma_A)
 \frac{1}{E+i\epsilon-\Heff}V\,.
\end{equation}
Comparing this expression with Eq.~\eqref{SdagS}, one sees that for $\Gamma_A=0$ the unitarity deficit provides the time-delay matrix $Q_\gamma$ for systems with uniform absorption \cite{savi03a}. The corresponding eigenvalues play an important role for the description of thermal emission from random media, as first discovered by Beenakker, who also obtained their multi-channel distribution at perfect coupling \cite{been98}. The relation to the time-delay matrix at finite absorption was later used to to generalize this result to the case of arbitrary coupling, $T\le1$ \cite{savi04}.

In the case of purely localized losses ($\epsilon=0$), Eq.~\eqref{S_ABS} also yields a useful pole-zero representation. In analogy with Eq.~\eqref{dets1}, one finds
\begin{equation}\label{detS_ABS}
 \det{S_A(E)} =\frac{\det{\left(E-{\cal H}_A\right)}}{\det{(E-\Heff)}},
 \qquad \mathcal{H}_A= H-\frac{i}{2}VV^{\dag}+\frac{i}{2}\Gamma_A
\end{equation}
Thus the poles of $S_A(E)$ are governed by $\Heff$, whereas its zeros are eigenvalues of the distinct non-Hermitian Hamiltonian $\mathcal{H}_A$. Real zeros of $\det S_A(E)$ correspond to incident coherent wavefronts that are completely absorbed by the cavity and therefore define the condition for coherent perfect absorption. For nonzero localized loss, part of the zero spectrum is now moved into the lower half-plane, and the density of such zeros determines the likelihood of realizing CPA in a complex enclosure. Within RMT this density was evaluated in several regimes in \cite{fyodorov2017CPA,osman2020chaotic}. The latter work also proposed that complex zeros of locally absorptive systems can be probed in standard scattering experiments with uniform absorption through the \emph{reflection time difference}, a certain generalization of the Wigner time delay; see \cite{osman2020chaotic} for details.

Subunitary scattering systems also admit a complex generalization of the Wigner time delay, where the resonant behavior in its real and imaginary parts is found to serve as a reliable indicator of the CPA condition \cite{Chen2021pre}. More generally, one may define a family of complex time delays in absorbing systems, based on Wigner, reflection, or transmission time delays. Remarkably, these quantities display a superuniversal statistics characterized by a robust algebraic $\tau^{-3}$ tail, independent of many system-specific details, which can be related to the topological properties of the corresponding singularities; see \cite{Shaibe2025} for further discussion and experimental verification.

\section{Conclusion}

The RMT approach reviewed in this chapter provides a unified framework for describing universal statistical features of open quantum or wave chaotic systems. Combined with non-perturbative techniques such as orthogonal and skew-orthogonal polynomials, biorthogonal expansions, Selberg-integral methods, integrable hierarchies, matrix integration and supersymmetry, it allows one to study such systems from both the ``outside'' (through fluctuations of scattering and transport observables) and the ``inside'' (through the statistics of resonances and resonance states). Its strength lies in reducing microscopic complexity to a small set of universal inputs: symmetry class, mean level spacing, number of channels, and coupling strengths. At the same time, RMT is intrinsically an ensemble theory and its application to individual systems relies on an ergodicity assumption, namely that spectral or parametric averages reproduce ensemble averages. Understanding when and how this happens requires a complementary semiclassical approach, in which scattering amplitudes are expressed as sums over classical trajectories. In this way, the equivalence between RMT and semiclassical predictions has been established for transport moments~\cite{muel09,berk12} and time-delay moments~\cite{Kuipers2014}, see further discussion and details in Chapters~\cite{Mueller_chapter} and~\cite{Novaes_chapter} of this volume.

We have not attempted to cover in detail the broad experimental landscape in which RMT predictions for chaotic scattering and transport have been tested. This is a substantial subject in its own right. Important experimental platforms include compound-nucleus reactions, where statistical scattering theory historically emerged \cite{mitc10}, mesoscopic quantum dots, where conductance and shot-noise fluctuations test the transport predictions of circular and Jacobi ensembles \cite{been97,alha00}, and electromagnetic systems, where wave propagation through complex media probes related universal phenomena \cite{shi15}. Microwave billiards and networks provide highly controllable realisations of wave-chaotic scattering and allow direct access to scattering matrices, resonance widths, impedance statistics, and absorption effects \cite{kuhl05a,grad14,kuhl13}. These experiments not only confirm many RMT predictions but also highlight the role of non-universal corrections, such as short trajectories, imperfect coupling,  and incomplete mode control; see the reviews mentioned above for further experimental perspectives.

On the theoretical side, several important developments go beyond the standard universal RMT regime considered here. One direction concerns systems with incomplete ergodicity, where eigenfunctions may be multifractal, localized, or only partially extended. Random band matrices, Rosenzweig--Porter type models, and related structured ensembles provide bridges between fully invariant RMT and spatially resolved disordered systems, allowing localization effects and non-ergodic extended phases to be incorporated into scattering and resonance statistics \cite{Fyo24,fyodmeib2025,DeTomasi2023}. Another direction concerns open disordered systems for which the separation into an internal chaotic region and external scattering channels is not natural, such as ensembles of randomly positioned scatterers or Euclidean random matrices \cite{Ski11,Goetschy2013}. Further developments involve genuinely non-Hermitian and sub-unitary random-matrix models~\cite{scho15}, motivated by gain/loss balance, coherent perfect absorption, and non-Hermitian topology; see \cite{Bergholtz2021,Ding2022} for recent reviews of related topological aspects. Closely related questions arise in many-body localization~\cite{Zakrzewski_chapter}, where intrinsically nonergodic many-body dynamics coupled to external continua or dissipative environments leads to new resonance and scattering problems. These directions suggest that the universal RMT picture of chaotic scattering remains a central reference point, while many current problems require more structured ensembles and refined notions of universality. This calls for a synthesis of RMT, semiclassics, supersymmetric field theory, and modern probabilistic approaches.

\begin{ack}[Acknowledgments]
\ The research at King's College London was supported by  EPSRC grant {\bf UKRI1015} "Non-Hermitian random matrices: theory and applications".
\end{ack}



\begin{thebibliography}{100}

\bibitem{verb85}
J.J.M.~Verbaarschot, H.A.~Weidenm{\"{u}}ller and M.R.~Zirnbauer,
  \emph{Grassmann integration in stochastical quantum physics: The case of
  compound-nucleus scattering},
  \href{https://doi.org/10.1016/0370-1573(85)90070-5}{\emph{Phys. Rep.}
  {\bfseries 129} (1985) 367}.

\bibitem{mitc10}
G.E.~Mitchell, A.~Richter and H.A.~Weidenm\"uller, \emph{Random matrices and
  chaos in nuclear physics: Nuclear reactions},
  \href{https://doi.org/10.1103/RevModPhys.82.2845}{\emph{Rev. Mod. Phys.}
  {\bfseries 82} (2010) 2845}.

\bibitem{been97}
C.W.J.~Beenakker, \emph{Random-matrix theory of quantum transport}, {\emph{Rev.
  Mod. Phys.} {\bfseries 69} (1997) 731}.

\bibitem{alha00}
Y.~Alhassid, \emph{The statistical theory of quantum dots}, {\emph{Rev. Mod.
  Phys.} {\bfseries 72} (2000) 895}.

\bibitem{shi15}
Z.~Shi, M.~Davy and A.Z.~Genack, \emph{Statistics and control of waves in
  disordered media}, \href{https://doi.org/10.1364/OE.23.012293}{\emph{Opt.
  Express} {\bfseries 23} (2015) 12293}.

\bibitem{kuhl05a}
U.~Kuhl, H.-J.~St{\"{o}}ckmann and R.~Weaver, \emph{Classical wave experiments
  on chaotic scattering},
  \href{https://doi.org/10.1088/0305-4470/38/49/001}{\emph{J. Phys. A}
  {\bfseries 38} (2005) 10433}.

\bibitem{mello_kumar2004book}
P.A.~Mello and N.~Kumar, \emph{Quantum Transport in Mesoscopic Systems:
  Complexity and Statistical Fluctuations. A Maximum Entropy Viewpoint}, Oxford
  University Press (2004).

\bibitem{guhr98}
T.~Guhr, A.~{M\"{u}ller-Groeling} and H.A.~Weidenm{\"{u}}ller, \emph{Random
  matrix theories in {Q}uantum {P}hysics: Common concepts},
  \href{https://doi.org/10.1016/S0370-1573(97)00088-4}{\emph{Phys. Rep.}
  {\bfseries 299} (1998) 189}.

\bibitem{soko89}
V.V.~Sokolov and V.G.~Zelevinsky, \emph{Dynamics and statistics of unstable
  quantum states},
  \href{https://doi.org/10.1016/0375-9474(89)90558-7}{\emph{Nucl. Phys. A}
  {\bfseries 504} (1989) 562}.

\bibitem{fyod97}
Y.V.~Fyodorov and H.-J.~Sommers, \emph{Statistics of resonance poles, phase
  shifts and time delays in quantum chaotic scattering: Random matrix approach
  for systems with broken time-reversal invariance},
  \href{https://doi.org/10.1063/1.531919}{\emph{J. Math. Phys.} {\bfseries 38}
  (1997) 1918}.

\bibitem{scho15}
H.~Schomerus, \emph{Random matrix approaches to open quantum systems},  in
  \emph{Stochastic Processes and Random Matrices: Lecture Notes of the Les
  Houches Summer School 2015}, G.S.~et~al, ed., pp.~409--473, Oxford University
  Press (2017).

\bibitem{fyod05r}
Y.V.~Fyodorov, D.V.~Savin and H.-J.~Sommers, \emph{Scattering, reflection and
  impedance of waves in chaotic and disordered systems with absorption},
  \href{https://doi.org/10.1088/0305-4470/38/49/017}{\emph{J. Phys. A}
  {\bfseries 38} (2005) 10731}.

\bibitem{Ski11}
S.E.~Skipetrov and A.~Goetschy, \emph{{Eigenvalue distributions of large
  Euclidean random matrices for waves in random media}}, {\emph{J. Phys. A
  Math. Theor.} {\bfseries 44} (2011) 65102}.

\bibitem{Livsic}
M.S.~Liv{\v{s}}ic, \emph{Operators, Oscillations, Waves (Open Systems), 
  Translations of Mathematical Monographs, Vol. 34}, American Mathematical
  Society, Providence, RI (1973).

\bibitem{fyod00}
Y.V.~Fyodorov and H.-J.~Sommers, \emph{Spectra of random contractions and
  scattering theory for discrete-time systems}, {\emph{JETP Lett.} {\bfseries
  72} (2000) 422}.

\bibitem{fyod11ox}
Y.V.~Fyodorov and D.V.~Savin, \emph{Resonance scattering of waves in chaotic
  systems},  in \emph{The Oxford Handbook of Random Matrix Theory}, G.~Akemann,
  J.~Baik and P.~Di~Francesco, eds., pp.~703--722, Oxford University Press, UK,
  2011 \href{https://doi.org/10.48550/arXiv.1003.0702}{[arXiv:1003.0702]}.

\bibitem{kott05}
T.~Kottos, \emph{Statistics of resonances and delay times in random media:
  Beyond random matrix theory}, {\emph{J. Phys. A} {\bfseries 38} (2005)
  10761}.

\bibitem{Gaspard2022}
D.~Gaspard and J.-M.~Sparenberg, \emph{Resonance distribution in the quantum
  random lorentz gas},
  \href{https://doi.org/10.1103/PhysRevA.105.042205}{\emph{Phys. Rev. A}
  {\bfseries 105} (2022) 042205}.

\bibitem{Fyo24}
Y.V.~Fyodorov, M.A.~Skvortsov and K.S.~Tikhonov, \emph{{Resonances in a
  single-lead reflection from a disordered medium: $\sigma$-model approach}},
  {\emph{Ann. Phys. (N. Y).} {\bfseries 460} (2024) 169568}.

\bibitem{fyodmeib2025}
Y.V.~Fyodorov and J.~Meibohm, \emph{Density of reflection resonances in
  one-dimensional disordered schr$\backslash$"$\{$o$\}$ dinger operators},
  \href{https://doi.org/10.1088/1367-2630/ae488d}{\emph{New Journal of Physics}
  {\bfseries 28} (2026) 034603}.

\bibitem{grad14}
G.~Gradoni, J.-H.~Yeh, B.~Xiao, T.M.~Antonsen, S.M.~Anlage and E.~Ott,
  \emph{Predicting the statistics of wave transport through chaotic cavities by
  the random coupling model: A review and recent progress},
  \href{https://doi.org/http://dx.doi.org/10.1016/j.wavemoti.2014.02.003}{\emph{Wave
  Motion} {\bfseries 51} (2014) 606}.

\bibitem{kuhl08}
U.~Kuhl, R.~H{\"{o}}hmann, J.~Main and H.-J.~St{\"{o}}ckmann, \emph{Resonance
  widths in open microwave cavities studied by harmonic inversion},
  \href{https://doi.org/10.1103/PhysRevLett.100.254101}{\emph{Phys. Rev. Lett.}
  {\bfseries 100} (2008) 254101}.

\bibitem{difa12}
A.~Di~Falco, T.F.~Krauss and A.~Fratalocchi, \emph{Lifetime statistics of
  quantum chaos studied by a multiscale analysis},
  \href{https://doi.org/10.1063/1.4711018}{\emph{Appl. Phys. Lett.} {\bfseries
  100} (2012) 184101}.

\bibitem{Kolovsky_chapter}
A.R.~Kolovsky, \emph{Chaotic dynamics and quantum transport},
  \href{https://doi.org/10.48550/arXiv.2605.12409}{\emph{Chapter in this
  volume; arXiv:2604.12409} (2026) }.

\bibitem{deCar02}
C.A.A.~de~Carvalho and H.M.~Nussenzveig, \emph{Time delay}, {\emph{Phys. Rep.}
  {\bfseries 364} (2002) 83}.

\bibitem{kolo13}
E.E.~Kolomeitsev and D.N.~Voskresensky, \emph{Time delays and advances in
  classical and quantum systems},
  \href{https://doi.org/10.1088/0954-3899/40/11/113101}{\emph{J. Phys. G: Nucl.
  Part. Phys.} {\bfseries 40} (2013) 113101}.

\bibitem{texier2016wigner}
C.~Texier, \emph{Wigner time delay and related concepts: Application to
  transport in coherent conductors}, {\emph{Physica E: Low-dimensional Systems
  and Nanostructures} {\bfseries 82} (2016) 16}.

\bibitem{soko97}
V.V.~Sokolov and V.~Zelevinsky, \emph{Simple mode on a highly excited
  background: Collective strength and damping in the continuum},
  \href{https://doi.org/10.1103/PhysRevC.56.311}{\emph{Phys. Rev. C} {\bfseries
  56} (1997) 311}.

\bibitem{lehm95b}
N.~Lehmann, D.V.~Savin, V.V.~Sokolov and H.-J.~Sommers, \emph{Time delay
  correlations in chaotic scattering: Random matrix approach}, {\emph{Physica
  D} {\bfseries 86} (1995) 572}.

\bibitem{Kieburg_chapter}
M.~Kieburg, \emph{Quantum chaotic systems: A random-matrix approach},
  \href{https://doi.org/10.48550/arXiv.2605.12141}{\emph{Chapter in this
  volume; arXiv:2604.12141} (2026) }.

\bibitem{lehm95a}
N.~Lehmann, D.~Saher, V.V.~Sokolov and H.-J.~Sommers, \emph{Chaotic scattering:
  The supersymmetry method for large number of channels}, {\emph{Nucl. Phys. A}
  {\bfseries 582} (1995) 223}.

\bibitem{brou95}
P.W.~Brouwer, \emph{Generalized circular ensemble of scattering matrices for a
  chaotic cavity with nonideal lead}, {\emph{Phys. Rev. B} {\bfseries 51}
  (1995) 16878}.

\bibitem{brou96}
P.W.~Brouwer and C.W.J.~Beenakker, \emph{Diagrammatic method of integration
  over the unitary group, with applications to quantum transport in mesoscopic
  systems}, {\emph{J. Math. Phys.} {\bfseries 37} (1996) 4904}.

\bibitem{gopa98}
V.A.~Gopar and P.A.~Mello, \emph{The problem of quantum chaotic scattering with
  direct processes reduced to one-without}, {\emph{Europhys. Lett.} {\bfseries
  42} (1998) 131}.

\bibitem{savi01}
D.V.~Savin, Y.V.~Fyodorov and H.-J.~Sommers, \emph{Reducing nonideal to ideal
  coupling in random matrix description of chaotic scattering: Application to
  the time-delay problem},
  \href{https://doi.org/10.1103/PhysRevE.63.035202}{\emph{Phys. Rev. E}
  {\bfseries 63} (2001) 035202(R)}.

\bibitem{Forrester}
P.~Forrester, \emph{Log-Gases and Random Matrices}, Princeton University Press
  (2010).

\bibitem{blan00r}
{\mbox{Ya}}.M.~Blanter and M.~B{\"{u}}ttiker, \emph{Shot noise in mesoscopic
  conductors}, {\emph{Phys. Rep.} {\bfseries 336} (2000) 1}.

\bibitem{arau98}
J.E.F.~Ara{\'{u}}jo and A.M.S.~Mac{\^{e}}do, \emph{Transport through quantum
  dots: A supersymmetry approach to transmission eigenvalue statistics},
  {\emph{Phys. Rev. B} {\bfseries 58} (1998) R13379}.

\bibitem{vivo08a}
P.~Vivo and E.~Vivo, \emph{Transmission eigenvalue densities and moments in
  chaotic cavities from random matrix theory}, {\emph{J. Phys. A} {\bfseries
  41} (2008) 122004}.

\bibitem{savi06}
D.V.~Savin and H.-J.~Sommers, \emph{Shot noise in chaotic cavities with an
  arbitrary number of open channels},
  \href{https://doi.org/10.1103/PhysRevB.73.081307}{\emph{Phys. Rev. B}
  {\bfseries 73} (2006) 081307(R)}.

\bibitem{savi08}
D.V.~Savin, H.-J.~Sommers and W.~Wieczorek, \emph{Nonlinear statistics of
  quantum transport in chaotic cavities},
  \href{https://doi.org/10.1103/PhysRevB.77.125332}{\emph{Phys. Rev. B}
  {\bfseries 77} (2008) 125332}.

\bibitem{nova07}
M.~Novaes, \emph{Full counting statistics of chaotic cavities with many open
  channels}, {\emph{Phys. Rev. B} {\bfseries 75} (2007) 073304}.

\bibitem{bulg06c}
E.N.~Bulgakov, V.A.~Gopar, P.A.~Mello and I.~Rotter, \emph{Statistical study of
  the conductunce and shot noise in open quantum chaotic cavities: Contribution
  from wispering galary modes}, {\emph{Phys. Rev. B} {\bfseries 73} (2006)
  155302}.

\bibitem{mezz11}
F.~Mezzadri and N.~Simm, \emph{Moments of the transmission eigenvalues, proper
  delay times, and random matrix theory. {I}},
  \href{https://doi.org/10.1063/1.3644378}{\emph{J. Math. Phys.} {\bfseries 52}
  (2011) 103511}.

\bibitem{mezz12}
F.~Mezzadri and N.~Simm, \emph{Moments of the transmission eigenvalues, proper
  delay times, and random matrix theory. {II}},
  \href{https://doi.org/10.1063/1.4708623}{\emph{J. Math. Phys.} {\bfseries 53}
  (2012) 053504}.

\bibitem{nova08}
M.~Novaes, \emph{Statistics of quantum transport in chaotic cavities with
  broken time-reversal symmetry}, {\emph{Phys. Rev. B} {\bfseries 78} (2008)
  035337}.

\bibitem{khor09}
B.A.~Khoruzhenko, D.V.~Savin and H.-J.~Sommers, \emph{Systematic approach to
  statistics of conductance and shot-noise in chaotic cavities}, {\emph{Phys.
  Rev. B} {\bfseries 80} (2009) 125301}.

\bibitem{Hua}
L.K.~Hua, \emph{Harmonic Analysis of Functions of Several Complex Variables in
  the Classical Domains}, American Mathematical Society, Providence, RI (1963).

\bibitem{osip08}
V.A.~Osipov and E.~Kanzieper, \emph{Integrable theory of quantum transport in
  chaotic cavities}, {\emph{Phys. Rev. Lett.} {\bfseries 101} (2008) 176804}.

\bibitem{osip09}
V.A.~Osipov and E.~Kanzieper, \emph{Statistics of thermal to shot noise
  crossover in chaotic cavities}, {\emph{J. Phys. A} {\bfseries 42} (2009)
  475101}.

\bibitem{mezz13}
F.~Mezzadri and N.~Simm, \emph{Tau-function theory of chaotic quantum transport
  with $\beta=1,2,4$},
  \href{https://doi.org/10.1007/s00220-013-1813-z}{\emph{Comm. Math. Phys.}
  {\bfseries 324} (2013) 465}.

\bibitem{Kumar2010}
S.~Kumar and A.~Pandey, \emph{Conductance distributions in chaotic mesoscopic
  cavities}, \href{https://doi.org/10.1088/1751-8113/43/28/285101}{\emph{J.
  Phys. A: Math. Theor.} {\bfseries 43} (2010) 285101}.

\bibitem{vivo08}
P.~Vivo, S.N.~Majumdar and O.~Bohigas, \emph{Distributions of conductance and
  shot noise and associated phase transitions}, {\emph{Phys. Rev. Lett.}
  {\bfseries 101} (2008) 216809}.

\bibitem{Cunden2015}
F.D.~Cunden, P.~Facchi and P.~Vivo, \emph{Joint statistics of quantum transport
  in chaotic cavities},
  \href{https://doi.org/10.1209/0295-5075/110/50002}{\emph{Europhys. Lett.}
  {\bfseries 110} (2015) 50002}.

\bibitem{Forrester2019}
P.J.~Forrester and S.~Kumar, \emph{Recursion scheme for the largest
  $\beta$-{W}ishart–{L}aguerre eigenvalue and {L}andauer conductance in
  quantum transport}, \href{https://doi.org/10.1088/1751-8121/ab433c}{\emph{J.
  Phys. A: Math. Theor.} {\bfseries 52} (2019) 42LT02}.

\bibitem{Vidal2012}
P.~Vidal and E.~Kanzieper, \emph{Statistics of reflection eigenvalues in
  chaotic cavities with nonideal leads},
  \href{https://doi.org/10.1103/PhysRevLett.108.206806}{\emph{Phys. Rev. Lett.}
  {\bfseries 108} (2012) 206806}.

\bibitem{Jarosz2015}
A.~Jarosz, P.~Vidal and E.~Kanzieper, \emph{Random matrix theory of quantum
  transport in chaotic cavities with nonideal leads},
  \href{https://doi.org/10.1103/PhysRevB.91.180203}{\emph{Phys. Rev. B}
  {\bfseries 91} (2015) 180203(R)}.

\bibitem{Rodriguez2013}
S.~Rodr\'{\i}guez-P\'erez, R.~Marino, M.~Novaes and P.~Vivo, \emph{Statistics
  of quantum transport in weakly nonideal chaotic cavities},
  \href{https://doi.org/10.1103/PhysRevE.88.052912}{\emph{Phys. Rev. E}
  {\bfseries 88} (2013) 052912}.

\bibitem{brou97a}
P.W.~Brouwer, K.M.~Frahm and C.W.J.~Beenakker, \emph{Quantum mechanical
  time-delay matrix in chaotic scattering}, {\emph{Phys. Rev. Lett.} {\bfseries
  78} (1997) 4737}.

\bibitem{brou99}
P.W.~Brouwer, K.M.~Frahm and C.W.J.~Beenakker, \emph{Distribution of the
  quantum mechanical time-delay matrix in chaotic cavity}, {\emph{Waves Random
  Media} {\bfseries 9} (1999) 91}.

\bibitem{Grabsch2018}
A.~Grabsch, D.V.~Savin and C.~Texier, \emph{Wigner-{S}mith time-delay matrix in
  chaotic cavities with non-ideal contacts},
  \href{https://doi.org/10.1088/1751-8121/aada43}{\emph{J. Phys. A: Math.
  Theor.} {\bfseries 51} (2018) 404001}.

\bibitem{fyod97a}
Y.V.~Fyodorov, D.V.~Savin and H.-J.~Sommers, \emph{Parametric correlations of
  phase shifts and statistics of time delays in quantum chaotic scattering:
  Crossover between unitary and orthogonal symmetries}, {\emph{Phys. Rev. E}
  {\bfseries 55} (1997) R4857}.

\bibitem{texi13}
C.~Texier and S.N.~Majumdar, \emph{Wigner time-delay distribution in chaotic
  cavities and freezing transition},
  \href{https://doi.org/10.1103/PhysRevLett.110.250602}{\emph{Phys. Rev. Lett.}
  {\bfseries 110} (2013) 250602}.

\bibitem{mart14}
A.M.~Mart\'inez-Arg\"uello, M.~Mart\'inez-Mares and J.C.~Garc\'ia, \emph{Joint
  moments of proper delay times},
  \href{https://doi.org/10.1063/1.4890559}{\emph{J. Math. Phys} {\bfseries 55}
  (2014) 081901}.

\bibitem{Kuipers2014}
J.~Kuipers, D.V.~Savin and M.~Sieber, \emph{Efficient semiclassical approach
  for time delays},
  \href{https://doi.org/10.1088/1367-2630/16/12/123018}{\emph{New J. Phys.}
  {\bfseries 16} (2014) 123018}.

\bibitem{Cunden2015prb}
F.D.~Cunden, \emph{Statistical distribution of the wigner-smith time-delay
  matrix moments for chaotic cavities},
  \href{https://doi.org/10.1103/PhysRevE.91.060102}{\emph{Phys. Rev. E}
  {\bfseries 91} (2015) 060102(R)}.

\bibitem{Cunden2016}
F.D.~Cunden, F.~Mezzadri, N.~Simm and P.~Vivo, \emph{Large-$n$ expansion for
  the time-delay matrix of ballistic chaotic cavities},
  \href{https://doi.org/10.1063/1.4966642}{\emph{J. Math. Phys.} {\bfseries 57}
  (2016) 111901}.

\bibitem{Novaes2015}
M.~Novaes, \emph{Statistics of time delay and scattering correlation functions
  in chaotic systems. i. random matrix theory},
  \href{https://doi.org/10.1063/1.4922746}{\emph{J. Math. Phys.} {\bfseries 56}
  (2015) 062110}.

\bibitem{Novaes2022}
M.~Novaes, \emph{Time delay statistics for finite number of channels in all
  symmetry classes},
  \href{https://doi.org/10.1209/0295-5075/ac806f}{\emph{Europhys. Lett.}
  {\bfseries 139} (2022) 21001}.

\bibitem{somm01}
H.-J.~Sommers, D.V.~Savin and V.V.~Sokolov, \emph{Distribution of proper delay
  times in quantum chaotic scattering: A crossover from ideal to weak
  coupling}, {\emph{Phys. Rev. Lett.} {\bfseries 87} (2001) 094101}.

\bibitem{savi03a}
D.V.~Savin and H.-J.~Sommers, \emph{Delay times and reflection in chaotic
  cavities with absorption},
  \href{https://doi.org/10.1103/PhysRevE.68.036211}{\emph{Phys. Rev. E}
  {\bfseries 68} (2003) 036211}.

\bibitem{Grabsch2020}
A.~Grabsch, \emph{Distribution of the {Wigner–Smith} time-delay matrix for
  chaotic cavities with absorption and coupled {Coulomb} gases},
  \href{https://doi.org/10.1088/1751-8121/ab58de}{\emph{J. Phys. A: Math.
  Theor.} {\bfseries 53} (2019) 025202}.

\bibitem{ditt00}
F.-M.~Dittes, \emph{The decay of quantum system with a small number of open
  channels}, {\emph{Phys. Rep.} {\bfseries 339} (2000) 215}.

\bibitem{diet09}
B.~Dietz, T.~Friedrich, H.L.~Harney, M.~Miski-Oglu, A.~Richter, F.~Sch{\"a}fer
  et~al., \emph{Induced violation of time-reversal invariance in the regime
  ofweakly overlapping resonances}, {\emph{Phys. Rev. Lett.} {\bfseries 103}
  (2009) 064101}.

\bibitem{gori05}
T.~Gorin, \emph{Random matrix description of decaying quantum systems},
  {\emph{J. Phys. A} {\bfseries 38} (2005) 10805}.

\bibitem{hagino2024microscopic}
K.~Hagino and G.F.~Bertsch, \emph{Microscopic derivation of transition-state
  theory for complex quantum systems}, {\emph{Journal of the Physical Society
  of Japan} {\bfseries 93} (2024) 064003}.

\bibitem{weidenmuller2024transition}
H.A.~Weidenm{\"u}ller, \emph{Transition-state theory reexamined},
  {\emph{Physical Review E} {\bfseries 109} (2024) 034117}.

\bibitem{ulla69}
N.~Ullah, \emph{On a generalized distribution of of the poles of the unitary
  collision matrix}, {\emph{J. Math. Phys.} {\bfseries 10} (1969) 2099}.

\bibitem{kozhan2017rank}
R.~Kozhan, \emph{Rank one non-hermitian perturbations of hermitian
  $\beta$-ensembles of random matrices}, {\emph{Journal of Statistical Physics}
  {\bfseries 168} (2017) 92}.

\bibitem{fyod99}
Y.V.~Fyodorov and B.A.~Khoruzhenko, \emph{Systematic analytical approach to
  correlation functions of resonances in quantum chaotic scattering},
  {\emph{Phys. Rev. Lett.} {\bfseries 83} (1999) 65}.

\bibitem{fyod03r}
Y.V.~Fyodorov and H.-J.~Sommers, \emph{Random matrices close to {H}ermitian or
  unitary: Overview of methods and results}, {\emph{J. Phys. A} {\bfseries 36}
  (2003) 3303}.

\bibitem{fyod_osm2022}
Y.V.~Fyodorov and M.~Osman, \emph{Eigenfunction non-orthogonality factors and
  the shape of cpa-like dips in a single-channel reflection from lossy chaotic
  cavities}, {\emph{Journal of Physics A: Mathematical and Theoretical}
  {\bfseries 55} (2022) 224013}.

\bibitem{somm99}
H.-J.~Sommers, Y.V.~Fyodorov and M.~Titov, \emph{{$S$}-matrix poles for chaotic
  quantum systems as eigenvalues of complex symmetric random matrices: From
  isolated to overlapping resonances}, {\emph{J. Phys. A} {\bfseries 32} (1999)
  L77}.

\bibitem{fyo2015res_pert}
Y.V.~Fyodorov and D.V.~Savin, \emph{Resonance width distribution in rmt:
  Weak-coupling regime beyond porter-thomas}, {\emph{Europhysics Letters}
  {\bfseries 110} (2015) 40006}.

\bibitem{kott00}
T.~Kottos and U.~Smilansky, \emph{Chaotic scattering on graphs}, {\emph{Phys.
  Rev. Lett.} {\bfseries 85} (2000) 968}.

\bibitem{chen2021stat_poles}
L.~Chen, S.M.~Anlage and Y.V.~Fyodorov, \emph{Statistics of complex wigner time
  delays as a counter of s-matrix poles: Theory and experiment},
  {\emph{Physical Review Letters} {\bfseries 127} (2021) 204101}.

\bibitem{Kur25}
M.S.~Kurilov and P.M.~Ostrovsky, \emph{{Density of scattering resonances in a
  disordered system}}, {\emph{arXiv:2512.18717} (2025) }.

\bibitem{dittes1991formation}
F.-M.~Dittes, H.~Harney and I.~Rotter, \emph{Formation of fast and slow decay
  modes in n-level systems coupled to one open channel}, {\emph{Physics Letters
  A} {\bfseries 153} (1991) 451}.

\bibitem{pers00}
E.~Persson, I.~Rotter, H.-J.~St{\"{o}}ckmann and M.~Barth, \emph{Observation of
  resonance trapping in open microwave cavity}, {\emph{Phys. Rev. Lett.}
  {\bfseries 85} (2000) 2478}.

\bibitem{dubach2023dynamics}
G.~Dubach and L.~Erd{\H{o}}s, \emph{Dynamics of a rank-one perturbation of a
  hermitian matrix}, {\emph{Electronic Communications in Probability}
  {\bfseries 28} (2023) 1}.

\bibitem{fyod_khor_popl_2022extreme}
Y.V.~Fyodorov, B.A.~Khoruzhenko and M.~Poplavskyi, \emph{Extreme eigenvalues
  and the emerging outlier in rank-one non-hermitian deformations of the
  gaussian unitary ensemble}, {\emph{Entropy} {\bfseries 25} (2022) 74}.

\bibitem{haak92}
F.~Haake, F.M.~Izrailev, N.~Lehmann, D.~Saher and H.-J.~Sommers,
  \emph{Statistics of complex levels of random matrices for decaying systems},
  {\emph{Z. Phys. B} {\bfseries 88} (1992) 359}.

\bibitem{akemann2025spectral}
G.~Akemann, Y.V.~Fyodorov and D.V.~Savin, \emph{Spectral density and
  eigenvector nonorthogonality in complex symmetric random matrices},
  {\emph{arXiv preprint arXiv:2511.21643} (2025) }.

\bibitem{mi2014Beenakker}
S.~Mi, D.~Pikulin, M.~Marciani and C.~Beenakker, \emph{X-shaped and y-shaped
  andreev resonance profiles in a superconducting quantum dot}, {\emph{Journal
  of Experimental and Theoretical Physics} {\bfseries 119} (2014) 1018}.

\bibitem{khoruzhenko2010truncations}
B.A.~Khoruzhenko, H.-J.~Sommers and K.~{\.Z}yczkowski, \emph{Truncations of
  random orthogonal matrices}, {\emph{Physical Review E—Statistical,
  Nonlinear, and Soft Matter Physics} {\bfseries 82} (2010) 040106}.

\bibitem{savi97}
D.V.~Savin and V.V.~Sokolov, \emph{Quantum versus classical decay laws in open
  chaotic systems},
  \href{https://doi.org/10.1103/PhysRevE.56.R4911}{\emph{Phys. Rev. E}
  {\bfseries 56} (1997) R4911}.

\bibitem{fyod_sav2012res_nonorth}
Y.V.~Fyodorov and D.V.~Savin, \emph{Statistics of resonance width shifts as a
  signature of eigenfunction nonorthogonality}, {\emph{Physical review letters}
  {\bfseries 108} (2012) 184101}.

\bibitem{schom00a}
H.~Schomerus, K.M.~Frahm, M.~Patra and C.W.J.~Beenakker, \emph{Quantum limit of
  the laser line width in chaotic cavities and statistics of residues of
  scattering matrix poles},
  \href{https://doi.org/10.1016/S0378-4371(99)00602-0}{\emph{Physica A}
  {\bfseries 278} (2000) 469}.

\bibitem{fyod02}
Y.V.~Fyodorov and B.~Mehlig, \emph{Statistics of resonances and nonorthogonal
  eigenfunctions in a model for single-channel chaotic scattering},
  {\emph{Phys. Rev. E} {\bfseries 66} (2002) 045202(R)}.

\bibitem{fyod12b}
Y.V.~Fyodorov and D.V.~Savin, \emph{Statistics of resonance width shifts as a
  signature of eigenfunction nonorthogonality},
  \href{https://doi.org/10.1103/PhysRevLett.108.184101}{\emph{Phys. Rev. Lett.}
  {\bfseries 108} (2012) 184101}.

\bibitem{poli09b}
C.~Poli, D.V.~Savin, O.~Legrand and F.~Mortessagne, \emph{Statistics of
  resonance states in open chaotic systems: A perturbative approach},
  \href{https://doi.org/10.1103/PhysRevE.80.046203}{\emph{Phys. Rev. E}
  {\bfseries 80} (2009) 046203}.

\bibitem{gros14}
J.-B.~Gros, U.~Kuhl, O.~Legrand, F.~Mortessagne, E.~Richalot and V.~Savin, D.\,
  \emph{Experimental width shift distribution: A test of nonorthogonality for
  local and global perturbations},
  \href{https://doi.org/10.1103/PhysRevLett.113.224101}{\emph{Phys. Rev. Lett.}
  {\bfseries 113} (2014) 224101}.

\bibitem{casa97}
G.~Casati, G.~Maspero and D.~Shepelyansky, \emph{Relaxation process in a regime
  of quantum chaos}, {\emph{Phys. Rev. E} {\bfseries 56} (1997) R6233}.

\bibitem{doro90}
E.~Doron, U.~Smilansky and A.~Frenkel, \emph{Experimental demonstration of
  chaotic scattering of microwaves}, {\emph{Phys. Rev. Lett.} {\bfseries 65}
  (1990) 3072}.

\bibitem{schaf03}
R.~Sch{\"{a}}fer, T.~Gorin, T.H.~Seligman and H.-J.~St{\"o}ckmann,
  \emph{Correlation functions of scattering matrix elements in microwave
  cavities with strong absorption},
  \href{https://doi.org/10.1088/0305-4470/36/12/325}{\emph{J. Phys. A}
  {\bfseries 36} (2003) 3289}.

\bibitem{savi06a}
D.V.~Savin, Y.V.~Fyodorov and H.-J.~Sommers, \emph{Correlation functions of
  impedance and scattering matrix elements in chaotic absorbing cavities},
  {\emph{Acta Phys. Pol. A} {\bfseries 109} (2006) 53}.

\bibitem{zhen06}
X.~Zheng, S.~Hemmady, T.M.~Antonsen, S.M.~Anlage and E.~Ott,
  \emph{Characterization of fluctuations of impedance and scattering matrix
  elements in wave chaotic scattering}, {\emph{Phys. Rev. E} {\bfseries 73}
  (2006) 046208}.

\bibitem{lawn10}
M.~Lawniczak, S.~Bauch, O.~Hul and L.~Sirko, \emph{Experimental investigation
  of the enhancement factor for microwave irregular networks with preserved and
  broken time reversal symmetry in the presence of absorption},
  \href{https://doi.org/10.1103/PhysRevE.81.046204}{\emph{Phys. Rev. E}
  {\bfseries 81} (2010) 046204}.

\bibitem{Bialous2023}
M.~Bia\l{}ous, B.~Dietz and L.~Sirko, \emph{Experimental study of the elastic
  enhancement factor in a three-dimensional wave-chaotic microwave resonator
  exhibiting strongly overlapping resonances},
  \href{https://doi.org/10.1103/PhysRevE.107.054210}{\emph{Phys. Rev. E}
  {\bfseries 107} (2023) 054210}.

\bibitem{hemm05}
S.~Hemmady, X.~Zheng, E.~Ott, T.M.~Antonsen and S.M.~Anlage, \emph{Universal
  impedance fluctuations in wave chaotic systems},
  \href{https://doi.org/10.1103/PhysRevLett.94.014102}{\emph{Phys. Rev. Lett.}
  {\bfseries 94} (2005) 014102}.

\bibitem{fyod04b}
Y.V.~Fyodorov and D.V.~Savin, \emph{Statistics of impedance, local density of
  states, and reflection in quantum chaotic systems with absorption},
  \href{https://doi.org/10.1134/1.1868794}{\emph{JETP Lett.} {\bfseries 80}
  (2004) 725}.

\bibitem{savi05}
D.V.~Savin, H.-J.~Sommers and Y.V.~Fyodorov, \emph{Universal statistics of the
  local green function in wave chaotic systems with absorption},
  \href{https://doi.org/10.1134/1.2150877}{\emph{JETP Lett.} {\bfseries 82}
  (2005) 544}.

\bibitem{kuhl05}
U.~Kuhl, M.~Mart\'{i}nez-Mares, R.A.~M\'{e}ndez-S\'{a}nchez and
  H.-J.~St{\"{o}}ckmann, \emph{Direct processes in chaotic microwave cavities
  in the presence of absorption},
  \href{https://doi.org/10.1103/PhysRevLett.94.144101}{\emph{Phys. Rev. Lett.}
  {\bfseries 94} (2005) 144101}.

\bibitem{Hemmady2012}
S.~Hemmady, T.M.~Antonsen, E.~Ott and S.M.~Anlage, \emph{Statistical prediction
  and measurement of induced voltages on components within complicated
  enclosures: A wave-chaotic approach},
  \href{https://doi.org/10.1109/TEMC.2011.2177270}{\emph{IEEE Trans.
  Electromagn. Compat.} {\bfseries 54} (2012) 758}.

\bibitem{lawn19}
M.~Lawniczak and L.~Sirko, \emph{Investigation of the diagonal elements of the
  {Wigner's} reaction matrix for networks with violated time reversal
  invariance}, \href{https://doi.org/10.1038/s41598-019-42123-y}{\emph{Sci.
  Rep.} {\bfseries 9} (2019) 5630}.

\bibitem{fede20}
S.B.~Fedeli and Y.V.~Fyodorov, \emph{Statistics of off-diagonal entries of
  wigner {$K$}-matrix for chaotic wave systems with absorption},
  \href{https://doi.org/10.1088/1751-8121/ab73ab}{\emph{J. Phys. A: Math.
  Theor.} {\bfseries 53} (2020) 165701}.

\bibitem{fyodorov2023APPA}
Y.~Fyodorov, \emph{On the density of complex eigenvalues of wigner reaction
  matrix in a disordered or chaotic system with absorption}, {\emph{Acta
  Physica Polonica A} {\bfseries 144} (2023) 447}.

\bibitem{lawniczak2020experimental}
M.~{\L}awniczak, B.~van Tiggelen and L.~Sirko, \emph{Experimental investigation
  of distributions of the off-diagonal elements of the scattering matrix and
  wigner's {$K$}-matrix for networks with broken time reversal invariance},
  {\emph{Physical Review E} {\bfseries 102} (2020) 052214}.

\bibitem{farooq2024coupled}
O.~Farooq, A.~Akhshani, M.~{\L}awniczak, M.~Bia{\l}ous and L.~Sirko,
  \emph{Coupled unidirectional chaotic microwave graphs}, {\emph{Physical
  Review E} {\bfseries 110} (2024) 014206}.

\bibitem{nock14}
A.~Nock, S.~Kumar, H.-J.~Sommers and T.~Guhr, \emph{Distributions of
  off-diagonal scattering matrix elements: Exact results},
  \href{https://doi.org/http://dx.doi.org/10.1016/j.aop.2013.11.006}{\emph{Ann.
  Phys.} {\bfseries 342} (2014) 103}.

\bibitem{kuma13}
S.~Kumar, A.~Nock, H.-J.~Sommers, T.~Guhr, B.~Dietz, M.~Miski-Oglu et~al.,
  \emph{Distribution of scattering matrix elements in quantum chaotic
  scattering},
  \href{https://doi.org/10.1103/PhysRevLett.111.030403}{\emph{Phys. Rev. Lett.}
  {\bfseries 111} (2013) 030403}.

\bibitem{kohnes2025exact}
S.~K{\"o}hnes, J.~Che, B.~Dietz and T.~Guhr, \emph{Exact results for the
  ericson transition in stochastic quantum scattering and experimental
  validation}, {\emph{arXiv preprint arXiv:2503.11516} (2025) }.

\bibitem{kohnes2026universality}
S.~K{\"o}hnes, J.~Che, B.~Dietz and T.~Guhr, \emph{A universality emerging in a
  universality: Derivation of the ericson transition in stochastic quantum
  scattering and experimental validation}, {\emph{arXiv preprint
  arXiv:2603.12068} (2026) }.

\bibitem{savi17}
D.V.~Savin, M.~Richter, U.~Kuhl, O.~Legrand and F.~Mortessagne,
  \emph{Fluctuations in an established transmission in the presence of a
  complex environment},
  \href{https://doi.org/10.1103/PhysRevE.96.032221}{\emph{Phys. Rev. E}
  {\bfseries 96} (2017) 032221}.

\bibitem{savi18}
D.V.~Savin, \emph{Envelope and phase distribution of a resonance transmission
  through a complex environment},
  \href{https://doi.org/10.1103/PhysRevE.97.062202}{\emph{Phys. Rev. E}
  {\bfseries 97} (2018) 062202}.

\bibitem{savi20}
D.V.~Savin, \emph{Statistics of a simple transmission mode on a lossy chaotic
  background},
  \href{https://doi.org/10.1103/PhysRevResearch.2.013246}{\emph{Phys. Rev.
  Research} {\bfseries 2} (2020) 013246}.

\bibitem{Wang2025}
C.-Z.~Wang, J.~Guillamon, U.~Kuhl, M.~Davy, M.~Reisner, A.~Goetschy et~al.,
  \emph{Optimal targeted mode transport in complex wave environments: A
  universal statistical framework},
  \href{https://doi.org/10.48550/arXiv.2501.12511}{\emph{arXiv:2501.12511}
  (2025) }.

\bibitem{fyodorov2023intensity}
Y.V.~Fyodorov and E.~Safonova, \emph{Intensity statistics inside an open
  wave-chaotic cavity with broken time-reversal invariance}, {\emph{Physical
  Review E} {\bfseries 108} (2023) 044206}.

\bibitem{brouwer2003wave}
P.W.~Brouwer, \emph{Wave function statistics in open chaotic billiards},
  {\emph{Physical Review E} {\bfseries 68} (2003) 046205}.

\bibitem{glue02}
M.~Gl{\"{u}}ck, A.R.~Kolovsky and H.J.~Korsch, \emph{Wannier--{S}tark
  resonances in optical and semiconductor superlattices},
  \href{https://doi.org/10.1016/S0370-1573(02)00142-4}{\emph{Phys. Rep.}
  {\bfseries 366} (2002) 103}.

\bibitem{jacq03}
{\mbox{Ph}}.~Jacquod, H.~Schomerus and C.W.J.~Beenakker, \emph{Quantum
  {Andreev} map: A paradigm of quantum chaos in superconductivity},
  {\emph{Phys. Rev. Lett.} {\bfseries 90} (2003) 207004}.

\bibitem{schom09}
H.~Schomerus, J.~Wiersig and J.~Main, \emph{Lifetime statistics in chaotic
  dielectric microresonators}, {\emph{Phys. Rev. A} {\bfseries 79} (2009)
  053806}.

\bibitem{nova13a}
M.~Novaes, \emph{Resonances in open quantum maps},
  \href{https://doi.org/10.1088/1751-8113/46/14/143001}{\emph{J. Phys. A}
  {\bfseries 46} (2013) 143001}.

\bibitem{zycz00}
K.~{\.{Z}}yczkowski and H.-J.~Sommers, \emph{Truncations of random unitary
  matrices}, {\emph{J. Phys. A} {\bfseries 33} (2000) 2045}.

\bibitem{signor2025beyond}
E.M.~Signor, M.A.P.~Reynoso, B.~Vijaywargia, S.D.~Prado and L.F.~Santos,
  \emph{Beyond ginibre statistics in open floquet chaotic systems with
  localized leaks}, {\emph{arXiv preprint arXiv:2512.09038} (2025) }.

\bibitem{rozh03}
I.~Rozhkov, Y.V.~Fyodorov and R.L.~Weaver, \emph{Statistics of transmitted
  power in multichannel dissipative ergodic structures},
  \href{https://doi.org/10.1103/PhysRevE.68.016204}{\emph{Phys. Rev. E}
  {\bfseries 68} (2003) 016204}.

\bibitem{savi06b}
D.V.~Savin, O.~Legrand and F.~Mortessagne, \emph{Inhomogeneous losses and
  complexness of wave functions in chaotic cavities},
  \href{https://doi.org/10.1209/epl/i2006-10358-3}{\emph{Europhys. Lett.}
  {\bfseries 76} (2006) 774}.

\bibitem{chong2010coherent}
Y.~Chong, L.~Ge, H.~Cao and A.D.~Stone, \emph{Coherent perfect absorbers:
  time-reversed lasers}, {\emph{Physical review letters} {\bfseries 105} (2010)
  053901}.

\bibitem{li2017random}
H.~Li, S.~Suwunnarat, R.~Fleischmann, H.~Schanz and T.~Kottos, \emph{Random
  matrix theory approach to chaotic coherent perfect absorbers},
  {\emph{Physical Review Letters} {\bfseries 118} (2017) 044101}.

\bibitem{fyodorov2017CPA}
Y.V.~Fyodorov, S.~Suwunnarat and T.~Kottos, \emph{Distribution of zeros of the
  s-matrix of chaotic cavities with localized losses and coherent perfect
  absorption: non-perturbative results}, {\emph{Journal of Physics A:
  Mathematical and Theoretical} {\bfseries 50} (2017) 30LT01}.

\bibitem{osman2020chaotic}
M.~Osman and Y.V.~Fyodorov, \emph{Chaotic scattering with localized losses:
  S-matrix zeros and reflection time difference for systems with broken
  time-reversal invariance}, {\emph{Physical Review E} {\bfseries 102} (2020)
  012202}.

\bibitem{been98}
C.W.J.~Beenakker, \emph{Thermal radiation and amplified spontaneous emission
  from a random medium}, {\emph{Phys. Rev. Lett.} {\bfseries 81} (1998) 1829}.

\bibitem{savi04}
D.V.~Savin and H.-J.~Sommers, \emph{Distribution of reflection eigenvalues in
  many-channel chaotic cavities with absorption}, {\emph{Phys. Rev. E}
  {\bfseries 69} (2004) 035201(R)}.

\bibitem{Chen2021pre}
L.~Chen, S.M.~Anlage and Y.V.~Fyodorov, \emph{Generalization of {W}igner time
  delay to subunitary scattering systems},
  \href{https://doi.org/10.1103/PhysRevE.103.L050203}{\emph{Phys. Rev. E}
  {\bfseries 103} (2021) L050203}.

\bibitem{Shaibe2025}
N.~Shaibe, J.M.~Erb and S.M.~Anlage, \emph{Superuniversal statistics of complex
  time delays in non-{H}ermitian scattering systems},
  \href{https://doi.org/10.1103/PhysRevLett.134.147203}{\emph{Phys. Rev. Lett.}
  {\bfseries 134} (2025) 147203}.

\bibitem{muel09}
S.~M{\"{u}}ller, S.~Heusler, A.~Altland, P.~Braun and F.~Haake,
  \emph{Periodic-orbit theory of universal level correlations in quantum
  chaos}, {\emph{New J. Phys.} {\bfseries 9} (2009) 103025}.

\bibitem{berk12}
G.~Berkolaiko and J.~Kuipers, \emph{Universality in chaotic quantum transport:
  The concordance between random matrix and semiclassical theories},
  \href{https://doi.org/10.1103/PhysRevE.85.045201}{\emph{Phys. Rev. E}
  {\bfseries 85} (2012) 045201}.

\bibitem{Mueller_chapter}
S.~M\"uller and M.~Sieber, \emph{Semiclassical periodic-orbit theory for
  quantum spectra},
  \href{https://doi.org/10.48550/arXiv.2605.19019}{\emph{Chapter in this
  volume; arXiv:2605.19019} (2026) }.

\bibitem{Novaes_chapter}
M.~Novaes, \emph{Semiclassical theory of transport},
  \href{https://doi.org/10.48550/arXiv.2604.13193}{\emph{Chapter in this
  volume; arXiv:2604.13193} (2026) }.

\bibitem{kuhl13}
U.~Kuhl, O.~Legrand and F.~Mortessagne, \emph{Microwave experiments using open
  chaotic cavities in the realm of the effective hamiltonian formalism},
  \href{https://doi.org/10.1002/prop.201200101}{\emph{Fortschr. Phys.}
  {\bfseries 61} (2013) 404}.

\bibitem{DeTomasi2023}
G.~De~Tomasi and I.M.~Khaymovich, \emph{Non-{H}ermiticity induces localization:
  Good and bad resonances in power-law random banded matrices},
  \href{https://doi.org/10.1103/PhysRevB.108.L180202}{\emph{Phys. Rev. B}
  {\bfseries 108} (2023) L180202}.

\bibitem{Goetschy2013}
A.~Goetschy and S.E.~Skipetrov, \emph{Euclidean random matrices and their
  applications in physics},
  \href{https://doi.org/10.48550/arxiv.1303.2880}{\emph{arXiv:1303.2880} (2013)
  }.

\bibitem{Bergholtz2021}
E.J.~Bergholtz, J.C.~Budich and F.K.~Kunst, \emph{Exceptional topology of
  non-hermitian systems},
  \href{https://doi.org/10.1103/RevModPhys.93.015005}{\emph{Rev. Mod. Phys.}
  {\bfseries 93} (2021) 015005}.

\bibitem{Ding2022}
K.~Ding, C.~Fang and G.~Ma, \emph{Non-hermitian topology and exceptional-point
  geometries}, \href{https://doi.org/10.1038/s42254-022-00516-5}{\emph{Nat.
  Rev. Phys.} {\bfseries 4} (2022) 745–760}.

\bibitem{Zakrzewski_chapter}
J.~Zakrzewski, \emph{Many-body localization},
  \href{https://doi.org/10.48550/arXiv.2604.12464}{\emph{Chapter in this
  volume; arXiv:2604.12464} (2026) }.

\end{thebibliography}

\providecommand{\href}[2]{#2}\begingroup\raggedright\endgroup

\end{document}